\documentclass[10pt]{article}
\usepackage[utf8]{inputenc}
\usepackage{amsmath,amssymb,graphicx}
\usepackage{hyperref}
\usepackage[T1]{fontenc}
\usepackage{mathpazo}
\usepackage[font=small,labelfont=bf]{caption}
\usepackage{longtable}
\usepackage{verbatim}
\usepackage{amsfonts}
\usepackage{cite}
\usepackage{float}
\usepackage[utf8]{inputenc}
\usepackage[section]{placeins}

\usepackage[table,usenames,dvipsnames]{xcolor}
\definecolor{darkerblue}{rgb}{0.2,0.2,0.5}

\usepackage{afterpage}
\usepackage[utf8]{inputenc}

\usepackage{wrapfig}
    
\usepackage{enumerate}

\newcommand{\bear}{\begin{array}}
\newcommand{\ear}{\end{array}}

\newcommand{\beq}{\begin{eqnarray}}
\newcommand{\eeq}{\end{eqnarray}}
\newcommand{\beqa}{\begin{eqnarray}}
\newcommand{\eeqa}{\end{eqnarray}}

\def\OMIT#1{{}}
\newcommand{\lsim}{\mathrel{\rlap{\lower4pt\hbox{\hskip1pt$\sim$}}
    \raise1pt\hbox{$<$}}}         
\newcommand{\gsim}{\mathrel{\rlap{\lower4pt\hbox{\hskip1pt$\sim$}}
    \raise1pt\hbox{$>$}}}         

\usepackage{tikz}
\usetikzlibrary{trees}
\usetikzlibrary{decorations.pathmorphing}
\usetikzlibrary{decorations.markings}
\tikzset{
    photon/.style={decorate, decoration={snake}, draw=black},
    wino/.style={draw=redwine},    
    electron/.style={draw=black, postaction={decorate},
        decoration={markings,mark=at position .55 with {\arrow[draw=black]{>}}}},
    scalar/.style={draw=black, dashed,postaction={decorate},
        decoration={markings,mark=at position .55 with {\arrow[draw=black]{>}}}},
    gluon/.style={decorate, draw=black,
        decoration={coil,amplitude=4pt, segment length=5pt}}
}

\usepackage{caption}
\usepackage{subcaption}
\usepackage{authblk}

\usepackage{titlesec}

\title{\bf \color{Blue} Randomness-Assisted Exponential Hierarchies}

\author[1]{Adam Tropper}
\author[1,2]{JiJi Fan}
\affil[1]{Department of Physics, Brown University, Providence, RI 02912}
\affil[2]{The Brown Theoretical Physics Center (BTPC), Providence, RI 02912}

\begin{document}
\maketitle

\begin{abstract}
Inspired by the localization phenomenon in condensed matter systems, we explore constructions in the theory space of multiple scalar fields, in which exponentially suppressed couplings could originate from random parameters. In particular, we find a new class of non-local theory space models, in which scalar fields at non-adjacent sites interact with each other but with strengths decaying exponentially with the site separation. Such a model could have very different localization properties, compared to the local theory space scenarios with only nearest-site interactions, based on the original Anderson localization model. More specifically, we find that a particular non-local interaction pattern leads to bi-localization of the two lightest eigenstates. Exponential localization (and thus exponentially suppressed couplings) then emerges only and immediately when randomness is introduced, no matter how tiny it is. We discuss variants of the model and possible UV completions as well. 
\end{abstract}


\section{Introduction}
\label{sec:intro}

In his seminal paper in 1958~\cite{PhysRev.109.1492}, Anderson showed that electron energy eigenstates localize in a single-particle system with short range hopping between adjacent sites, even in the presence of disorder. Since then, localization and phase transitions between extended ergodic and localized phases have been hot topics in condensed matter physics. Over the last 60 years, there have been significant numerical and analytical efforts exploring Anderson localization and generalizations of Anderson's original model. For some recent reviews, see~\cite{abrahams201050, Izrailev_2012}. One interesting direction to extend the original Anderson model is to consider single-particle systems with long-range hopping, that is, hopping between non-adjacent sites on a lattice. Examples include consider a constant hopping between any arbitrary pair of sites~\cite{Owusu_2008, Ossipov_2013, Modak_2016, Celardo_2016} and long-range hopping that follows a power-law decay~\cite{1989EL......9...83L, PhysRevLett.64.547, Mirlin_1996, Evers_2008, 1989JETPL..50..338B, Deng_2018, Nosov_2019}. One of the most important findings is that the correlation in the long-range hopping tends to localize the system~\cite{Nosov_2019}. 

The localization phenomenon in condensed matter systems could have interesting analogues in particle physics. Recently, it is pointed out in ref.~\cite{Craig:2017ppp} that one might apply a new mechanism analogous to the original Anderson model in one dimension, in which mass eigenstates along a local theory space localize with random mass parameters, to generate exponential hierarchies in four-dimensional quantum field theories.\footnote{The popular ``clockwork" model, originally proposed in refs~\cite{Choi:2014rja, Choi:2015fiu, Kaplan:2015fuy} also bear some similarities to the Anderson model in the sense that they also include nearest-site interactions, but they have no random parameters.} Other analogies have also been discussed in the context of inflation~\cite{Green:2014xqa}, particle production in the early Universe~\cite{Amin:2015ftc} and high-dimensional gravitational theory~\cite{Rothstein:2012hk}. 

In this article, we first introduce and study a new class of random Hamiltonians with translational-invariant long-range hopping that decays exponentially with separation between any two sites. We will demonstrate that unlike the models with power-law long-range hopping in the condensed matter literature, the energy eigenstates are exponentially localized in this case. 

It may be difficult to realize this Hamiltonian in a condensed matter system, in which it is more natural to expect that the Coloumb or dipole interactions lead to power-law hopping. Yet it is plausible to realize such kind of exponentially-decaying long-range hopping using quantum field theory in a non-local theory space. A theory space~\cite{ArkaniHamed:2001ca} consists of a set of sites denoting global or gauge symmetry groups, connected by links -- quantum fields that transform under adjacent groups. In this paper, we consider two toy models of $N$ real scalars in the theory space with exponentially decaying non-local interactions (i.e., interactions between scalars at non-adjacent sites, not to be confused with interactions that traverse the real space), that could be generated at loop levels through tree-level interactions between adjacent scalars. They both result in similar scalar mass matrices and their mass eigenstates are exponentially localized. Yet the details of the localization, such as how localization emerges, localization length and whether they could be UV completed, could be quite different. In particular, one type of construction and its variants could strengthen localization of the lightest mass eigenstates by making them more peaked at one site in the theory space, in the limit of small random diagonal perturbations. This example addresses one of the major motivations of our study: to explore and identify novel classes of theory space models based on random matrices with different localization properties, especially those that could improve localization.

Exponential localization leads to exponentially suppressed couplings between different sites in the theory space and thus could be used to generate exponential hierarchies. As is known, there exist quite a few puzzling hierarchies in fundamental physics, including the minuscule cosmological constant, the hierarchy problem of a light elementary Higgs boson, and the standard model fermion mass hierarchy. Any new mechanism that could potentially explain a large hierarchy is worth exploring, even when its early versions may not fully solve the problems above. The main interesting feature of the localization-inspired field theory constructions is that the localization emerges only when randomness is introduced. This is in sharp contrast to most of the other approaches to hierarchies, which rely on either continuous or discrete symmetries. In the local theory space model~\cite{Craig:2017ppp}, one still needs the random diagonal masses to be of order the nearest-site interactions for the exponential localization to be evident. 
In this article, in the non-local theory space model with the improved localization, localization at a single site originates from bi-localization due to the model structure without any randomness and then emerges immediately when one turns on even a tiny random perturbation.

The paper is organized as follows: in Sec.~\ref{sec:localization}, we review the basics of localization phenomena in single-particle systems and random matrix theory, the main tool to study the localization property. In Sec.~\ref{sec:hamiltonian}, we present the new class of Hamiltonians with random on-site energies and exponentially decaying long-range hopping. We demonstrate the localization of eigenstates through both numerical and analytical studies. In Sec.~\ref{sec:anderson-inspired}, we review the local theory space construction, inspired by the original Anderson model. In Sec.~\ref{sec:nonlocaltheoryspacemodels}, we introduce and study two non-local theory space toy models of $N$ real scalars, analyze their exponential localization properties and discuss their feasibilities from the UV point of view as well as the localization property of a variant model. We conclude in Sec.~\ref{sec:conclusion}.

\section{Localization in Condensed Matter Systems}
\label{sec:localization}

We begin this section by defining localization in random matrix theory and introducing the simplest random matrix model with exponential localization in the context of the famous one-dimensional Anderson model. 
We also discuss several important generalizations including long-range interactions. This section is mostly a review of basic concepts, tools, and models in the condensed matter literature, which are most relevant to our studies and can be safely skipped by readers who are familiar with the subject. 

\subsection{Definition of Localization}
\label{sec:defn of localization}
The mathematical tools used to study Anderson localization are random matrix theory, a subject analyzing the statistical properties of various ensembles of matrices with some or all of their elements pulled from prescribed probability distributions. The statistical properties that have been analyzed to great success include eigenvalue distributions, level spacing, repulsion of eigenvalues, norms of matrix powers, and typical eigenvector properties. For reviews, see~\cite{furstenberg1960products, mehta2004random, borot2017introduction, scardicchio2017perturbation}. 

The statistical property that we are most interested in is the existence of a ``localized" phase - i.e. a region of parameter space where eigenvectors are typically localized \cite{Nosov_2019}.\footnote{Note that when we claim that $X$ is a typical property of a random matrix ensemble, it means that $X$ holds in general except, perhaps, on a probability zero subset of the chosen distribution for the random parameters.} Consider an $N \times N$ random matrix model with $N$ being a free parameter. We say that the eigenvectors of the model are \textit{localized} if their support sets are independent of $N$. In contrast, \textit{fully ergodic} states have support on essentially all components while \textit{weakly ergodic} states have support for approximately $cN$ components where $0<c<1$. In between the localized and ergodic phases, there may exist a \textit{fractal} phase characterized by eigenvectors with support sets scaling like $N^D$ with $0<D<1$ \cite{Mirlin_1996, Nosov_2019}.

One way to assess numerically whether an eigenvector $v$ of an $N \times N$ random matrix is localized is to compute its $q$-th moment, defined by
\begin{equation}
    I_q(v) \equiv \sum_{j=1}^N |v_{j}|^{2q},
\label{def moments}
\end{equation}
where $v_j$ gives the $j^{th}$ component of the vector~\cite{Wegner1980, Mirlin_1993}. It has been argued that
\begin{equation}
    I_q(v) \propto \begin{cases}
\xi^{-(q-1)} &\text{if~} v \text{~is localized}\\
N^{-D(q-1)}&\text{if~}v \text{~is fractal} \\
(2q-1)!! \, N^{-(q-1)}&\text{if~}v \text{~is ergodic},
\end{cases}
\label{eq:moments cases}
\end{equation}
where $\xi$ is the size of the support set~\cite{Wegner1980, Castellani_1986, Mirlin_1996}.

One could also determine the existence of different phases using the simple Mott and Levitov conditions~\cite{mott1966electrical, PhysRevLett.64.547}. These criteria can be applied to any matrix model $M$ with entries taking the form $M_{j,k}= \epsilon_{j}\delta_{j,k} + m_{j,k}$, where $\epsilon_j$'s are random parameters with $\langle \epsilon_j \rangle = 0$ and $\langle \epsilon_j^2 \rangle = \sigma^2\neq 0$. $m$ could be deterministic or random. Mott's criterion holds for eigenstates in the bulk of the spectrum and is given by 
\begin{equation}
    \lim_{N \rightarrow \infty} \frac{\sigma}{|\Delta(m)|} = 0 \hspace{10pt} \Longrightarrow \hspace{10pt} \text{(at least weak) ergodicity },
\label{Mott Criteria}
\end{equation}
where $\Delta(m)$ is the entire set of eigenvalues of the matrix $m$, usually referred to as the \textit{spectrum} of $m$. The difference between largest and smallest eigenvalues of $m$, $|\Delta(m)| = \max (\Delta(m)) - \min (\Delta(m))$, is the \textit{bandwidth} of the spectrum \cite{Nosov_2019, mott1966electrical}. On the other hand, Levitov's criterion holds for all eigenvectors and is given by
\cite{Nosov_2019, PhysRevLett.64.547}:
\begin{equation}
    \lim_{N \rightarrow \infty} \frac{1}{N|\Delta(m)|} \sum_{j \neq k} \langle | m_{jk} | \rangle < \infty \hspace{10pt} \Longrightarrow \hspace{10pt} \text{localization}.
\label{Levitov Criteria}
\end{equation}
Note that both criteria are sufficient but not necessary conditions.

There are also different classes of localized states. An eigenstate of a matrix model is \textit{exponentially localized} if it can be well-approximated by an exponentially decaying envelope, that is, if the components of the eigenvector take the form 
\begin{equation}
    |v_j| \sim e^{-\frac{|j-j_0|}{L}},
\label{localization length}
\end{equation} 
where $j_0$ labels the component of $v$ with the largest absolute value and $L$ is a constant that is referred to as the \textit{localization length} of the eigenvector. $L$ characterizes the spread of the eigenstate around $j_0$. Similarly, we say that an eigenvector is \textit{algebraically localized} if its components decay as an algebraic function. In the rest of the article, we will refer to localization as being synonymous with exponential localization unless stated otherwise. While one can use numerical simulations to show that components of an eigenvector are well-fit by some function for a finite $N$, it is often important to demonstrate that the fitting is a genuine feature of the model and not just some numerical artifact at small $N$'s by showing that eigenvector components converge to the function arbitrarily far away from $j_0$. To this end, one often uses a locator expansion method~\cite{PhysRev.109.1492, Pichard_1986, hundertmark2008short,scardicchio2017perturbation} (more discussions will be presented in section \ref{sec:locatorexpansion}) or a transfer matrix approach~\cite{thouless1972relation,herbert1971localized}. 

\subsection{Anderson Localization}
\label{sec: Anderson Localization}
The simplest random matrix model with exponentially localized eigenstates was studied by Anderson in the pioneering paper~\cite{PhysRev.109.1492}. The model contains a Hamiltonian of a single particle moving on a lattice of $N$ sites:
\begin{equation}
    \left(H_A\right)_{j,k} = \epsilon_j \delta_{j,k} + g(\delta_{j,k+1}+\delta_{j+1,k}),
 \label{eq:anderson}
\end{equation}
where the subscripts $j$ and $k$ denote the sites, $g$ is a coupling constant dictating the strength of the nearest-neighbor hopping and $\epsilon_j$'s are random parameters pulled from a uniform distribution in the interval $[0,W]$ which corresponds to local potential defects on the lattice. It is argued that there exists a critical $W_c\geq0$ such that for all $W> W_c$, eigenstates of the Hamiltonian are exponentially localized with probability one~\cite{PhysRev.109.1492}. A crucial result is that in one dimension, $W_c = 0$, which was proven rigorously using transfer matrices~\cite{thouless1972relation,herbert1971localized} and the locator expansion technique~\cite{PhysRev.109.1492, Pichard_1986, hundertmark2008short,scardicchio2017perturbation}.

The Anderson model has two interesting limits: strong and weak localization regimes, where the average eigenvector localization lengths are short and long respectively. In the strong localization regime, $W \gg g$. Localization is unsurprising since the Hamiltonian becomes nearly diagonal. In this regime, localization length is given by \cite{Izrailev_2012} 
\begin{equation}
    L(g,W) \sim \left(\ln \frac{W}{g} -1 \right)^{-1}.
\label{eqn: W >> g}
\end{equation}
In the weak localization regime, $W \ll g$, the localization length becomes eigenvalue dependent (we denote the eigenvalue $\lambda$), and is given by \cite{Izrailev_2012}:
\begin{equation}
    L(\lambda,g,W) \sim \begin{cases} 26.3 \big(\frac{g}{W}\big)^2 & \lambda \sim 0, \\
    5 \big(\frac{g}{W}\big)^{\frac{2}{3}} & \big(2 - \frac{|\lambda|}{g}\big)\big(\frac{3g^2}{W^2}\big)^{\frac{2}{3}} \ll 1,  \\
    24\big(\frac{g}{W}\big)^2 -6 \big(\frac{\lambda}{W}\big)^2 & \text{otherwise}.
    \end{cases}
\label{eqn: W << g}
\end{equation}
While localization in this regime is, indeed, much weaker, eigenstates with eigenvalues near the edges of the spectrum $\Delta(H) \subset [-2g,2g+W]$ exhibit stronger localization relative to bulk eigenstates. In addition, it is known that if one chooses to vary $g$ between adjacent sites, the model still typically has exponentially localized eigenvectors provided that the pairs $(\epsilon_j,g_j)$ are chosen independently from a two-dimensional distribution \cite{theodorou1976extended}.

Another important discovery is that the Anderson Hamiltonian can be generalized to a square lattice in two dimensions with local defects and nearest neighbor hopping. These models also feature exponential localization of eigenstates. But for $d \geq 3$, there is only localization when $W > W_c$ with $W_c > 0$. Furthermore, $W_c$ is a function of eigenvalues in this case. Again the eigenstates near the band edges are more localized with smaller values of $W_c$~\cite{scardicchio2017perturbation}.

\subsection{Non-local Models} 
\label{sec:non-local models}
The Anderson model only includes nearest-neighbor hopping interactions. Since then, there has been a growing interest in studying variants of the original model adding non-local interactions.

One simple example with long-range hopping is the Yuzbashyan-Shastry Hamiltonian for a one-dimensional lattice \cite{Modak_2016}
\begin{equation}
    H_{j,k} = \epsilon_{j} \delta_{j,k} + N^{-\gamma/2}, \hspace{30pt} \text{with} \hspace{30pt} \langle \epsilon_j \rangle = 0~,~\langle \epsilon_j^2 \rangle = \sigma^2.
\end{equation} 
In this case, the interactions are \textit{fully correlated} since all the off-diagonal matrix elements are identical~\cite{Nosov_2019}. It has been shown that there are no truly extended states for all $\sigma$ and $\gamma$'s, which is quite surprising because the model has an approximate cyclic symmetry, which is broken only by the diagonal perturbations. Indeed, all Hamiltonians with cyclic symmetries are necessarily characterized by completely delocalized Bloch states~\cite{Nosov_2019}. The localization is interpreted as an effect of destructive interference of long-range hopping trajectories in the presence of a (full) correlation between the long-range interactions.

Instead of having a fully correlated model, one could also consider a \textit{partially correlated} model with Hamiltonian 
\begin{equation}
    H_{j,k} = \epsilon_{j} \delta_{j,k} + t_{j-k} N^{-\gamma/2}, \hspace{30pt} \text{with} \hspace{30pt} \langle \epsilon_j \rangle = \langle t_{j-k} \rangle = 0~;~\langle \epsilon_j^2 \rangle = \langle |t_{j-k}|^2\rangle = \sigma^2,
\end{equation} 
in which the hopping terms still preserves the translational invariance. A \textit{totally uncorrelated} model is given by the Rosenzweig-Porter Hamiltonian~\cite{rosenzweig1960repulsion} 
\begin{equation}
    H_{j,k} = \epsilon_{j} \delta_{j,k} + t_{j,k} N^{-\gamma/2}, \hspace{30pt} \text{with} \hspace{30pt} \langle \epsilon_j \rangle = \langle t_{j,k} \rangle = 0~;~\langle \epsilon_j^2 \rangle = \langle |t_{j,k}|^2\rangle = \sigma^2,
\end{equation}
 where there is no symmetry left and $t_{j,k}$'s only need to satisfy the Hermicity requirement  $t_{j,k}^* = t_{k,j}$.\footnote{The difference between fully correlated, partially correlated, and uncorrelated models has to do with the size of the symmetry group of the Hamiltonian in the absence of the $\epsilon_j$ terms. The fully correlated Hamiltonian is invariant under transformations in the symmetric group, $S_n$, while the partially correlated Hamiltonian is invariant under transformations in the cyclic group $Z_n \subset S_n$, and the uncorrelated Hamiltonian is not invariant under any non-trivial transformation in any subgroup of $S_n$.} These two ensembles still have a localized phase, but only for $\gamma > 2$ \cite{Nosov_2019}. When $\gamma < 2$,  the two models are in either the fractal or ergodic states. While it may not appear surprising that localization exists for matrices with $\gamma > 2$ at least in the large $N$ limit with all the off-diagonal terms approaching zero, it is not obvious why the systems are delocalized when $0 < \gamma < 2$. The three examples above demonstrate that long-range interactions do not necessarily delocalize the system. However, increasing correlations in the non-local hopping terms generally leads to improved localization, a phenomenon dubbed \textit{correlation induced localization} in ref.~\cite{Nosov_2019}.

In general, translationally invariant Hamiltonians are particularly interesting since they arise often in physical situations. It could be useful to apply the Fourier transformation and study the system in the momentum space. More specifically, a Hamiltonian of the form $H_{j,k} = \epsilon_{j}\delta_{j,k} + h_{j-k}$ can be Fourier transformed to the dual one $\tilde{H}_{p,q} = \tilde{E}_p\delta_{p,q} + \tilde{J}_{p-q}$ with
\begin{equation}
    \tilde{E}_p = \sum_{j=0}^{N-1} h_j e^{-2 \pi i \frac{p j}{N}}, \hspace{30pt} \tilde{J}_{p} = \frac{1}{N} \sum_{j=0}^{N-1} \epsilon_{j} e^{-2 \pi i \frac{p j}{N}},
\end{equation}
assuming periodic boundary conditions~\cite{Nosov_2019}. Note that in the momentum space, the diagonal random entries and the off-diagonal interactions in the real space switch their roles. Thus it is easy to diagonalize the translation-invariant models in the momentum space when one sets aside the random diagonal perturbations. 
The Fourier transform is particularly useful for models with Hamiltonians parametrized by the family of power-law banded matrices: $H_{j,k} = \epsilon_{j}\delta_{j,k} + [1+(|j-k|/b)^{a}]^{-1}$, where $a$ and $b$ are positive adjustable free parameters. These models with algebraically decaying off-diagonal couplings have algebraically (rather than exponentially) localized eigenvectors, $v_j \sim |j-j_0|^{-a}$~\cite{Mirlin_1996}.

\section{Hamiltonian with Long-range Exponentially Decaying Hopping}
\label{sec:hamiltonian}
In this section, we introduce a new class of Hamiltonian with long-range interactions for a particle moving on a one-dimensional lattice with $N$ sites: 
\beq
\left(H_{\rm long-range}\right)_{j, k} = \epsilon_j \delta_{j,k} + \frac{g}{b^{|j-k|}} \left(1-\delta_{j,k} \right),
\label{eq:nonlocalH2}
\eeq
where both $\epsilon$'s and $g$ have the dimension of energy and $b$ is dimensionless. We take $b>1$. Analogous to the Anderson model, $\epsilon_j$'s are random parameters pulled from a uniform distribution in the interval $[0,W]$. In this model, the long-range hopping decays as an exponential function of the site separation. The hopping terms are translationally invariant, which means that the hopping between two different sites only depends on their separation but not on their specific positions. Here we only consider fully deterministic hopping, but one could generalize the discussions to random hopping with varying $g$. 

\subsection{Numerical Results}
\label{sec: Hamiltonian Numerics}

\begin{figure}[h]
    \centering
    \includegraphics[width = 1.0\textwidth]{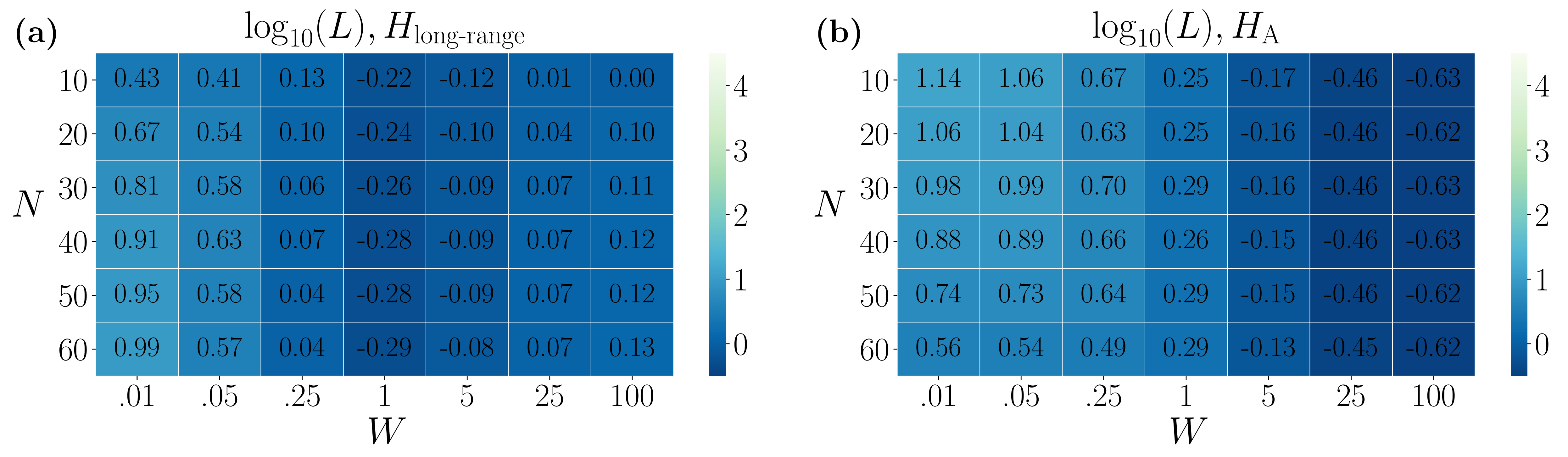}
    \caption{Median localization lengths for the Hamiltonian ensembles with exponentially-decaying long-range interactions (left) and the Anderson model (right). We fix $g = 1/2$ in the Anderson model in Eq.~\eqref{eq:anderson} and $g = 1$ and $b = 2$ in the long-range interacting model in Eq.~\eqref{eq:nonlocalH2}.}
    \label{fig:Hamiltonian Heatmaps}
\end{figure}

We first present the numerical properties, i.e., localization lengths, of the Hamiltonian ensembles defined in Eq.~\eqref{eq:nonlocalH2}, taking as an ansatz that they are characterized by exponentially localized eigenvectors. Indeed we find that numerically, the eigenvectors are well-fit by exponential envelopes. In the next section, we will also provide analytical evidence for the exponential localization. We generate a sample of 300 random matrices for given $N$ and $W$, take the lowest energy eigenvector for each matrix in the sample and compute its localization length by fitting it with an exponential envelope. We thus obtain a distribution for the localization length. The median localization lengths (more precisely, their logarithms) as functions of $N$ and $W$, for both the model with exponentially-decaying long-range interaction and the Anderson model, are shown in the heat maps in Figure~\ref{fig:Hamiltonian Heatmaps}.\footnote{We choose to present the median localization length rather than the mean localization length in anticipation for our quantum field theoretic models where there appear to be some small probability sets wherein the localization length is many orders of magnitude bigger than the median localization length - skewing the statistics in a misleading way.} The darker the region is, the shorter the localization length is and the more localized the lightest eigenstate is. In the plot, we have also fixed $g=1/2$ for the Anderson model and $g = 1$, $b = 2$ for the exponential long-range model so that their nearest neighbor hopping terms are the same. Note that $g$ and $\epsilon$'s are in the same unit of energy, which we don't specify and does not affect the discussions.

From Figure~\ref{fig:Hamiltonian Heatmaps}, we confirm numerically that exponentially-decaying, long-range interactions do not harm exponential localization in contrast to other long-range interactions. It is not entirely surprising since the long-range hopping decays exponentially and becomes negligible when the hopping distance is large. In addition, in the $W \ll g$ limit, the localization improves slightly in the long-range model in the sense that the localization length shrinks a bit. In the opposite limit with $W \gg g$, however, the localization worsens in the long-range model.

\subsection{Analytical Proof of Exponential Localization}

\subsubsection{Localization Criteria}
We first apply the Mott's and Levitov's criteria reviewed in Sec.~\ref{sec:defn of localization} to determine the localization phase of our model. To use the criteria, we need to evaluate the bandwidth of the spectrum. The hopping terms could be represented by a Toeplitz matrix, $t$, which is translationally invariant and satisfies $t_{j,k} \equiv t_{k-j}$. Spectral properties of Toeplitz matrices have been studied extensively in the math literature, and we review the relevant mathematical tools and apply them to our model in Appendix~\ref{sec: bounded spectra (appendix)}. We only quote the final result below. As long as $b>1$, we have the spectrum in the full range
\beq
\Delta(H) \subset \left[-\frac{2}{b+1}g, \; \frac{2}{b-1}g+W \right], 
\label{eq:deltaH}
\eeq
so, the bandwidth is $|\Delta(H)| = \frac{4 b }{b^2-1}g + W$. The variance of the random terms in the model is $\langle \epsilon^2 \rangle \equiv \sigma^2 = W^2/3$.
Thus the Mott's criterion for ergodic states in Eq.~\eqref{Mott Criteria} is never satisfied. Since Mott's criterion is a sufficient but necessary condition, its violation cannot guarantee localization. On the other hand, Levitov's criterion in Eq.~\eqref{Levitov Criteria} is always satisfied since
\begin{equation}
    \lim_{N \rightarrow \infty} \frac{1}{N|\Delta(H)| } \sum_{j \neq k} \langle |H_{j,k}| \rangle = \lim_{N \rightarrow \infty} \frac{g}{N |\Delta(H)| } \sum_{j \neq k} \frac{1}{b^{|j-k|}} < \infty,
\end{equation}
where $\sum_{j \neq k} \frac{1}{b^{|j-k|}} \propto N$. Since Levitov's criterion is a sufficient condition of localization, we prove that the eigenstates of our model are localized. Yet, it does not tell us the detailed properties of the localization, which we will study in the next section.

\subsubsection{Analytical Approach in the Strong Localization Regime}
\label{sec:locatorexpansion}
Now we use the locator expansion method and forward scattering approximation to show that localization is exponential and to estimate the localization length in the strong localization regime. While the emergence of localization is trivial in the $W=\infty$ limit, it is still non-trivial to confirm that this localization phase is stable when including hopping with strength $g/W \ll 1$. More importantly, we want to know whether the localization follows an exponential or power law and what the localization length is.  
We will first review the methodology in the context of the Anderson model and then modify and adjust the derivation to our model. For complete treatments see refs.~\cite{PhysRev.109.1492, Pichard_1986, hundertmark2008short, scardicchio2017perturbation}, which offer more mathematical insights concerning the general theory. 

~

\noindent {\it Review of the Method.} In the Anderson model, in the infinite diagonal disorder limit, we could write the Hamiltonian as
\begin{equation}
    H_A = H_0 + gT, \quad {\rm with} \quad H_0 = \sum_i \epsilon_i |i\rangle \langle i|,  \quad T = \sum_{ i,j=i+1 } \left( |i\rangle \langle j| + |j \rangle \langle i| \right),
\end{equation}
where $i, j$ label the sites on the lattice. As discussed before, $\epsilon$'s are taken randomly from a uniform distribution in the range $[0, W]$. 
Consider the resolvent defined by
\beq
G(E) = \frac{1}{E-H}, \quad \forall E \notin \Delta(H),
\eeq
where $\Delta(H)$ is the spectrum of $H$. We denote the set of eigenstates of $H$ as $\{ \psi_\alpha\}$. 
The matrix element of the resolvent in the site basis is given by 
\begin{equation}
    G(i,j,E) = \sum_{\alpha, \beta} \langle i | \psi_\alpha \rangle \langle \psi_\alpha \left | \frac{1}{E-H} \right |\psi_\beta \rangle \langle \psi_\beta | j\rangle = \sum_{\alpha} \frac{\psi^*_\alpha(j) \psi_\alpha(i)}{E-E_\alpha},
\end{equation}
which immediately implies
\begin{equation}
    \psi^*_\alpha(j)\psi_\alpha(i) = \lim_{E \rightarrow E_\alpha} (E-E_\alpha)G(i,j,E). 
\end{equation}
In the limit of large diagonal randomness, the size of an eigenvector $\psi_\alpha$ at site $n$ with its main support located at site $a$ (so that $\psi_\alpha^*(a) \approx 1$) is given by
\begin{equation}
    \psi_\alpha(n) = \lim_{E \rightarrow \epsilon_a} (E-\epsilon_a)G(n,a,E).
    \label{eq:eigenvectoratn}
\end{equation}
The eigenvector is exponentially localized if $\psi_a(n)$ decays exponentially in the large $n$ limit. 

Given the general discussion above, it is clear that our goal is to calculate the resolvent, which we could carry it out in perturbation theory with the expansion parameter $g/W$. The unperturbed resolvent is $G_0(E) = (E-H_0)^{-1}$ and $G_0(i,j,E) = \delta_{i,j}/(E-\epsilon_i)$. Using the matrix equality $A^{-1} - B^{-1} = B^{-1}(B-A)A^{-1}$ with $A = E-H$ and $B = E-H_0$, we get $G(E) = G_0(E) + gG_0TG$. We iterate this equation and find
\begin{equation}
    G = G_0 + \sum_{n=1}^\infty g^n(G_0T)^nG_0.
\end{equation}
We could rewrite this series summation in a random-walk representation:
\begin{equation}
    G(i,j,E) = \frac{\delta_{i,j}}{E-\epsilon_i} + \frac{1}{E-\epsilon_i} \sum_{p: i \rightarrow j} g^{|p|} \prod_{s=1}^{|p|} \frac{1}{E-\epsilon_{p(s)}},
    \label{eq:randomwalk}
\end{equation}
where $p: i \rightarrow j$ is a path starting from the $i$-th site and ending at the $j$-th site with a total length $|p|$. $p(s)$ labels the site at the $s$-th step in the path. This expression could be further refined as a sum of self-avoiding walks by factorizing all closed loops out of the summation. Combining Eq.~\eqref{eq:eigenvectoratn} and~\eqref{eq:randomwalk}, we have
\begin{equation}
    \psi_\alpha(n) =  \sum_{p: a \rightarrow n} g^{|p|} \prod_{s = 1}^{|p|} \frac{1}{\epsilon_a-\epsilon_{p(s)}}. 
\end{equation}

To simplify the computation, we proceed with the forward scattering approximation. In this approximation, we only consider the maximal term in the sum, which corresponds to the shortest path as all longer paths are higher order in $g/W$. For simplicity, we set $\alpha=a=0$ and $\epsilon_a = 0$. This gives
\begin{equation}
    \frac{1}{n} \ln \left|\psi_0(n)\right| \sim  \ln(g) - \frac{1}{n} \sum_{s=1}^n \ln(\epsilon_{p(s)}) =  -\ln\frac{W}{g}+ 1
\end{equation}
where we use the central limit theorem and replace $\sum_{s=1}^n \ln(\epsilon_{p(s)})$ by $n \langle \ln(\epsilon_{p(s)}) \rangle =  \frac{n}{W} \int_0^W \ln(\epsilon) d \epsilon = n \left(\ln(W)-1\right)$. This shows that in the strong localization limit, $\psi$ falls off exponentially away from the dominant supporting site: $\left|\psi_0(n)\right| \propto e^{-n/L}$ with the localization length given by
\begin{equation}
L \sim \left(-\lim_{n \rightarrow \infty} \frac{1}{n} \ln|\psi_0(n)| \right)^{-1} = \left[\ln(W/g) -1\right]^{-1},
\end{equation}
which is positive and finite in the large randomness limit, $W \gg g$.

~

\noindent {\it Application to our model.} We are now ready to apply a similar strategy to study localization in the large randomness regime of our model. In our model, the perturbation Hamiltonian changes to
\beq
T =  \sum_{ i,j\neq i } \frac{1}{b^{|i-j|}}  \left( |i\rangle \langle j| + |j \rangle \langle i| \right). 
\eeq
The random walk representation of the resolvent is 
\begin{equation}
    G(i,j,E) = \frac{\delta_{i,j}}{E-\epsilon_i} + \frac{1}{E-\epsilon_i} \sum_{p:~i \rightarrow j} g^{|p|} \prod_{s=1}^{|p|}\left(b^{-|p(s)-p(s-1)|}\right)\frac{1}{E-\epsilon_{p(s)}}. 
\end{equation}
Comparing the equation above with the representation of Anderson model in Eq.~\eqref{eq:randomwalk}, one could see that the main difference is that the coupling for each step of random walk could vary, depending on the step length. To use Eq.~\eqref{eq:eigenvectoratn} to determine the shape of the eigenfunction $\psi_0$ from $G(0,n,E)$, we consider two possible paths: a one-step path $p_1: 0 \rightarrow n$ and a $n$-step path: $p_n: 0 \rightarrow 1 \rightarrow 2 \cdots \rightarrow n$. Taking into account only one of the two paths at a time, we find that
\beqa
p_n: && \frac{1}{n} \ln \left|\psi_0(n)\right| \sim   -\ln\frac{W}{g}+ 1 - \ln b , \nonumber \\
p_1: && \frac{1}{n} \ln \left|\psi_0(n)\right| \sim   - \ln b. 
\eeqa
In the limit $W/g \gg 1$, the wavefunction at site $n$ receives its dominant contribution from $p_1$. The other paths' contributions are also suppressed compared to that of $p_1$. The estimate above is crude but suffices to demonstrate the existence of exponential localization in the large randomness limit. Taking into account $p_1$ only, we estimate the localization length to be
\beq
L \sim \left(\ln b\right)^{-1},
\eeq
which we confirm to be consistent with numerical results.

Let's compare our model with the power-law long-range interaction model, in which the off-diagonal hopping scales as $|i-j|^{-\beta}$. Considering a single hop from the $0$-th to the $n$-th site (the $p_1$ path discussed above), we find that
\begin{equation}
    \psi_0(n) \sim - \frac{1}{n^\beta} \frac{1}{ \epsilon_n}. 
\end{equation}
The wave function decays algebraically as $n^{-\beta}$, consistent with results in the literature, e.g., ref.~\cite{Nosov_2019}. 

These techniques can also be used to estimate the localization length for the theory space models that we will discuss in the ensuing sections. It is not difficult to check that in the large randomness limit, the localization length in the theory space models are the same as the ones of the corresponding Hamiltonian models.

\section{Localization in Theory Space I: Anderson-Inspired Models}
\label{sec:anderson-inspired}

Now we discuss how the localization ideas from condensed matter systems with Hamiltonians represented by random matrices can be adapted to construct quantum field theory (QFT) models. In this and next section, we consider toy QFT models consisting of a large number of real scalars $\pi$'s. In other words, we consider the theory space of scalar fields with each scalar field corresponding to a site on the one-dimensional lattice reviewed in section~\ref{sec:localization}. Instead of studying the Hamiltonian, we study the mass matrix of the scalars, $M_\pi$. 

The Anderson localization inspired construction has already been considered in refs~\cite{Craig:2017ppp}, which we will briefly review below. Consider a model of $N$ real scalars $\pi_i$ with the following Lagrangian:
\begin{equation}
    \mathcal{L}_{\rm A} = \frac{1}{2} \sum_{j=1}^N (\partial_\mu \pi_j)^2 - \frac{1}{2} \sum_{j=1}^N \epsilon_j \pi_j^2 - \frac{1}{2} \sum_{i=1}^{N-1} g(\pi_j - \pi_{j+1})^2,
\label{Craig-Sutherland Lagrangian}
\end{equation} 
where $\epsilon_i$ are drawn randomly from a uniform distribution over $[0,W]$. This toy model could arise from a model of complex scalar fields, $\Phi_i$'s,\footnote{This complex scalar model still needs to be UV completed, which we will not discuss further.}
\begin{equation}
  \mathcal{L}_0=  \sum_{j=1}^N |\partial_\mu \Phi_j|^2 - V(\Phi_j^\dagger \Phi_j) - \left(\frac{1}{4} \sum_{j=1}^N\epsilon_j \Phi_j^2 + \sum_{j=1}^{N-1} g \Phi_{j}^\dagger \Phi_{j+1} + \text{h.c.}\right). 
\label{eqn: C+S UV Completion}
\end{equation}
The complex scalar UV completion has a $U(1)^N$ symmetry, which we take to be spontaneously broken by $V(\Phi^\dagger \Phi)$ and explicitly broken by the rest of the potential terms. The spontaneous breaking gives rise to a set of goldstone bosons $\pi$'s, which are the phase modes of the $\Phi$'s. $\pi$'s obtain a potential with the quadratic terms given in Eq.~\eqref{Craig-Sutherland Lagrangian} from the explicit breaking terms. 

This model is equivalent to: 
\begin{equation}
    \mathcal{L}_{\rm A}^\prime = \frac{1}{2} \sum_{j=1}^N (\partial_\mu \pi_j)^2 - \frac{1}{2} \sum_{j=1}^N \epsilon_j \pi_j^2 - \frac{1}{2} \sum_{i=1}^{N-1} g(\pi_j + \pi_{j+1})^2,
\label{Craig-Sutherland Lagrangian (Plus)}
\end{equation} 
which could be obtained from Eq.~\eqref{Craig-Sutherland Lagrangian} by a field redefinition $\pi_{\text{even}} \mapsto - \pi_{\text{even}}$ while keeping $\pi_{\text{odd}}$ unchanged. Thus their localization properties are identical.

The Lagrangians in Eq.~\eqref{Craig-Sutherland Lagrangian} and Eq.~\eqref{Craig-Sutherland Lagrangian (Plus)} clearly resemble the Anderson model. The real scalar $\pi_i$ corresponds to the $i$-th site on the 1d lattice. The diagonal mass term with coefficient $\epsilon_i$ plays the role of the on-site energy while the mass mixing terms correspond to nearest neighbor hopping. In the strong localization limit, when $W \gg g$, each mass eigenstate obviously has support dominantly from only one of the real scalars. In the more interesting weak localization limit, when $W \ll g$ and the diagonal mass terms are perturbations, the lightest eigenstate is more localized compared to the other eigenstates, analogous to the states at band edges in the Anderson model. Note that the nearest neighbor mixing terms only break down $U(1)^N$ to $U(1)$ while $\epsilon$ terms break $U(1)^N$ entirely. Thus, in the weak localization limit, the lightest eigenstate remains as a massless Goldstone boson before adding the diagonal perturbations. 

Numerically the localization length is quite large in the weak localization limit with large $g/W$, which is the more interesting limit compared to the strong localization limit from the model building point of view. For example, $g/W \sim 20$, one needs to include at least $N \sim 30$ scalar fields to achieve an exponential hierarchy $e^{-N/L}$ of order 0.2 (see panel (b) in Fig.~\ref{fig:LagrangianHeatmaps}). Relatively weak localization persists even when $g/W$ is reduced to be of order 1. Thus it will be interesting to explore other theory space models based on random matrices, which could have different localization properties and potentially improve the localization quality with $g/W \gtrsim 1$.

\section{Localization in the Theory Space II: Non-Local Theory Space Models}
\label{sec:nonlocaltheoryspacemodels}

In this section, we study non-local theory space models, which are inspired by the Hamiltonian model introduced in Section~\ref{sec:hamiltonian}. We first consider two toy models for $N$ real scalars $\pi$'s with Lagrangian $\mathcal{L}_+$ and $\mathcal{L}_-$: 
\begin{equation}
    \mathcal{L}_{\pm} = \frac{1}{2} \sum_{i=1}^N (\partial_\mu \pi_{i})^2 - \frac{1}{2} \sum_{j=1}^N \epsilon_j \pi_j^2 - \frac{1}{2} \sum_{i=1}^{N-1} \sum_{j=i+1}^{N} \frac{g}{b^{j-i}} (\pi_i \pm \pi_j)^2,
\label{eq:non-Local Lagrangian1}
\end{equation} 
where $\epsilon_j$ is pulled from the uniform distribution $[0,W]$. The $\epsilon_j$'s and $g$ have mass dimension two, and $b > 1$ is dimensionless.
$\mathcal{L}_+$ and $\mathcal{L}_-$ only differ by relative signs in the second potential term. While these two models appear very similar, and, indeed, all of them have the lightest eigenstate exponentially localized at large but finite $N$, they differ significantly in both embedding into UV completions and localization properties such as localization lengths for a given set of parameters. On the contrary, in the local theory space model reviewed in the previous section, we could change $(\pi_j - \pi_{j+1})^2$ to $(\pi_j + \pi_{j+1})^2$ without changing any localization property and both models have very similar UV completions. We will start by examining the localization properties of these two relatively simple models in subsection \ref{sec: L pm for finite N} and then explain differences in the UV completion of these models in subsection \ref{sec:UV completions}. We choose to proceed in this order because subsection \ref{sec:UV completions} introduces a third Lagrangian which is motivated solely from an interesting UV completion but is most easily studied and understood once we have already gained perspective on the toy models. The localization properties of the third Lagrangian are discussed in our final subsection, \ref{sec:Lmixed}.

\subsection{Localization of Toy Models in the Finite $N$ Limit}
 \label{sec: L pm for finite N}
 
In this section, we will present both numerical and analytical results of the localization properties of $\mathcal{L}_\pm$ in Eq.~\eqref{eq:non-Local Lagrangian1}. As we shall discuss in the next section, $\mathcal{L}_+$ does not have a simple perturbative UV completion while $\mathcal{L}_-$ does. Nevertheless, we will address both of them here as they have qualitatively different localization properties, which helps understand more complicated UV completions.

\subsubsection{Numerical Results}
\label{Lagrangian Numerics}

\begin{figure}[h]
    \centering
   \includegraphics[width=1.0\textwidth]{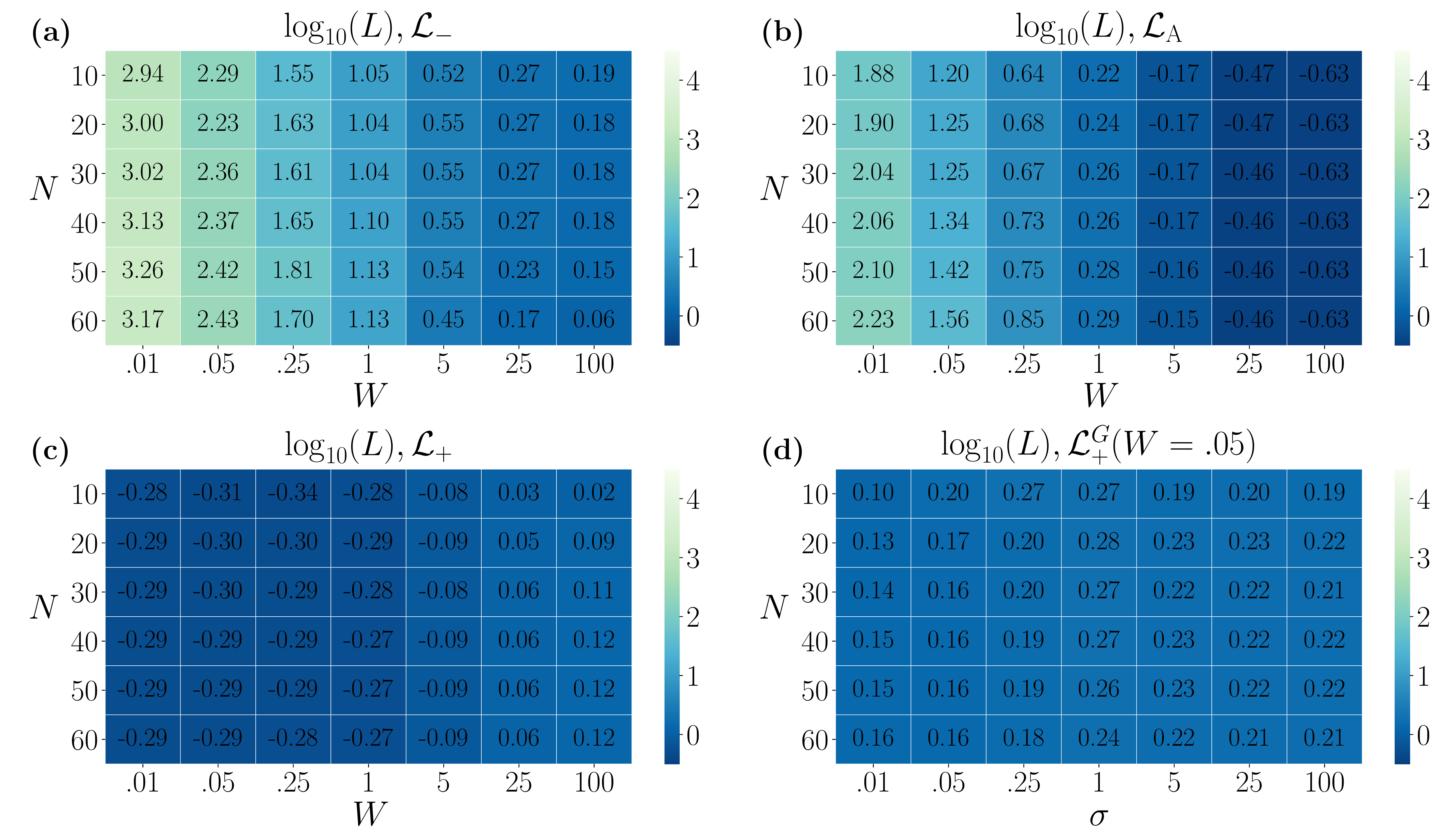}
    \caption{Median localization lengths of various toy models for different choices of parameters. The first three panels are the median localization lengths of ${\mathcal L}_-, {\mathcal L}_{\rm A}$ and ${\mathcal L}_+$ for given $(W, N)$ pairs. The panel on the bottom right is median localization lengths of $\mathcal{L}_+^G$ for given $(\sigma, N)$ pairs, with $W = 0.05$ and $g_{j-i}$ takes from a Gaussian distribution $f (x| \, 1,\sigma)$.}
    \label{fig:LagrangianHeatmaps}
\end{figure}

\begin{figure} [h]
    \centering
    \includegraphics[width=1.0\textwidth]{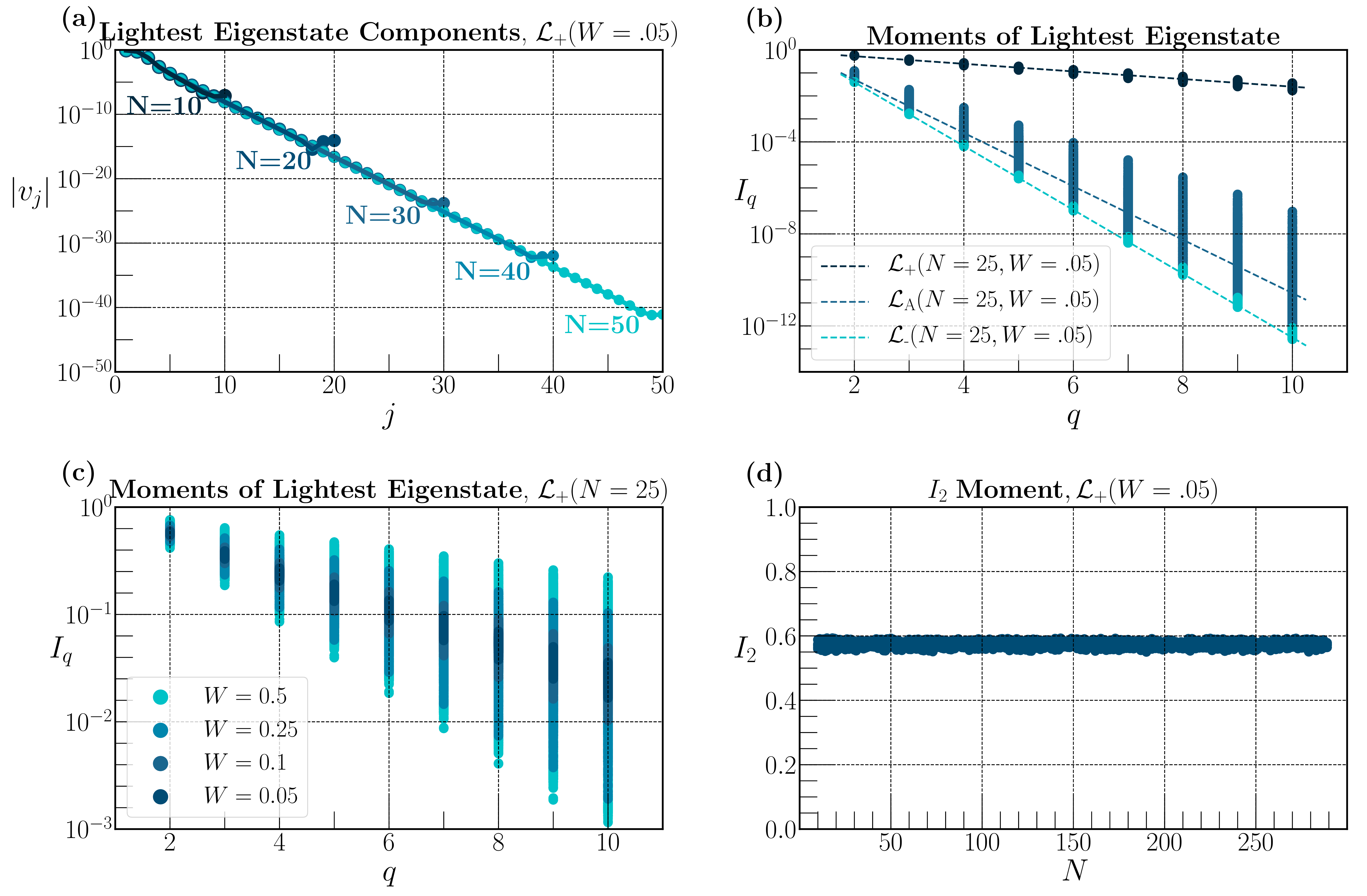}
    \caption{Panel (a): Absolute values of the lightest eigenvector components for different choices of $N$ in the $\mathcal{L}_+$ model. Panel (b): Moments $I_q$, defined in Eq. \eqref{def moments}, of the lightest eigenvectors, as a function of $q$ for three different models. The dashed lines join the average $I_q$ at each $q$. Panel (c): $I_q$ of the lightest eigenstates in the $\mathcal{L}_+$ model for different choices of $W$. Panel (d): $I_2$ of the lightest eigenstates in the $\mathcal{L}_+$ model as a function of $N$.}
    \label{fig:moments}
\end{figure}

We compute the localization lengths of the lightest mass eigenstates for three different types of random matrix ensembles based on ${\cal L}_{\rm A}$ and ${\cal L}_\pm$ at different $N$'s and $W$'s and the median of the results are presented in the first three panels of Figure~\ref{fig:LagrangianHeatmaps}. From now on, we fix $g=1/2$ in ${\cal L}_{\rm A}$ and $(g, b) = (1,2)$ in ${\cal L}_\pm$, where all the quantities are in the same (arbitrary) energy units. By this choice, the mass mixing between nearest neighbors takes the same value $1/2$ in all three models. 
We see two interesting effects:
\begin{itemize}
\item The inclusion of the non-local interactions in $\mathcal{L}_{-}$ strictly {\it worsens} the localization of $\mathcal{L}_{\text{A}}$ for all pairs of $(N,W)$ studied. 
\item The inclusion of the non-local interactions in $\mathcal{L}_{+}$ strictly {\it enhances} the localization of $\mathcal{L}_{\text{A}}$ in the small disorder regime with $g/W \gg 1$. 
\end{itemize}
Indeed, as we will explain in the following section, for arbitrary small diagonal mass perturbations ($W \ll g$), the lightest mass eigenstates in the $\mathcal{L}_+$ model typically have a localization length $L \sim \mathcal{O}(1)$. In addition, an $\mathcal{O}(1)$ localization length is present even in the intermediate region $W \sim g$, indicating that strong localization of the lightest eigenstate is attained throughout the entire parameter space. Note that these properties of $\mathcal{L}_+$ persist with the mean localization length as well as the median. As briefly commented in section \ref{sec: Hamiltonian Numerics}, the reason we choose to report the median rather than the mean is due to the fact that the distribution of the localization length could have a small tail extending to very high values for some random matrix ensembles. Yet for the $\mathcal{L}_+$ model, there is no such tail in the distribution of $L$. In other words, there are no set of random choices of diagonal mass perturbations that ruin the localization accidentally.

We also want to comment that the nice localization properties of $\mathcal{L}_+$ are robust even when we further allow random variations in the non-local mass mixings.
Consider $\mathcal{L}_+^G$ defined as follows:
\begin{equation}
    \mathcal{L}_{+}^G = \frac{1}{2} \sum_{i=1}^N (\partial_\mu \pi_{i})^2 - \frac{1}{2} \sum_{j=1}^N \epsilon_j \pi_j^2 - \frac{1}{2} \sum_{i=1}^{N-1} \sum_{j=i+1}^{N} \frac{g_{j-i}}{b^{j-i}} (\pi_i + \pi_j)^2,
\label{eq:gaussian lagrangian}
\end{equation} 
where the superscript $G$ stands for Gaussian and indicates that $g_{j-i}$ should be interpreted as parameters pulled independently from a normalized Gaussian distribution centering around $g=1$ with a variance $\sigma$, $f(x| \, 1, \sigma)$. In this case, the off-diagonal terms in the mass matrix are only partially correlated. The median localization length is presented in the last panel of Figure \ref{fig:LagrangianHeatmaps}. As expected, adding randomness in the non-local mass mixing increases the localization length, e.g., by a factor of 3 when increasing $\sigma=0$ to $\sigma=0.01$, and thus weakens localization. Yet an $\mathcal{O}(1)$ localization length is still maintained in the $\mathcal{L}_{+}^G$ model even when $\sigma$ is allowed to be as large as $\sim 100$, demonstrating that the strong localization properties of $\mathcal{L}_+$ are preserved under this generalization. There is a subtly: allowing $g_{j-i}$ to be pulled from a Gaussian distribution means that these parameters can now assume negative values. This implies that the mass-matrix for $\mathcal{L}_+^G$ may not be always positive-definite and there could be tachyonic directions in the field space. Accordingly, it is necessary to introduce positive quartic terms of $\pi$'s to stabilize their potential for this model.

Another way to study the exponential localization of ${\mathcal L}_+$ is to use moments of the eigenvectors defined in Eq.~\eqref{eq:moments cases}. The results are demonstrated in Fig.~\ref{fig:moments}. In panel (a), we show the components of lightest eigenstates for different choices of $N$. It is clear that the lightest eigenstates are exponentially localized. In panel (b), we compute and present $I_q$'s of the lightest eigenstates from ensembles of mass matrices with $N=25, W=0.05$ based on three toy models. 
Since all three models are localized, $I_q(v) \propto \xi^{-(q-1)}$ according to Eq.~\eqref{eq:moments cases}, where $\xi$, the size of the support set, is related to the localization length and also measures how localized the state is: the smaller $\xi$ is, the more localized the state is. From the panel, we could see that $\xi ({\mathcal L}_+) < \xi({\mathcal L}_{\rm A}) < \xi ({\mathcal L}_-)$, corroborating our findings using the median localization lengths. In panel (c), we show $I_q$'s for different $W$'s in the ${\mathcal L}_+$ model. The larger $W$ is, the more spread $I_q$ is for a given $q$, suggesting a weakened localization. Lastly, in panel (d), we present $I_2$ as a function of $N$, fixing $W=0.05$. Indeed $I_q$ is independent of $N$, confirming again that the lightest eigenstate is localized. 

\begin{figure}[H]
    \centering
    \includegraphics[width=1.0\textwidth]{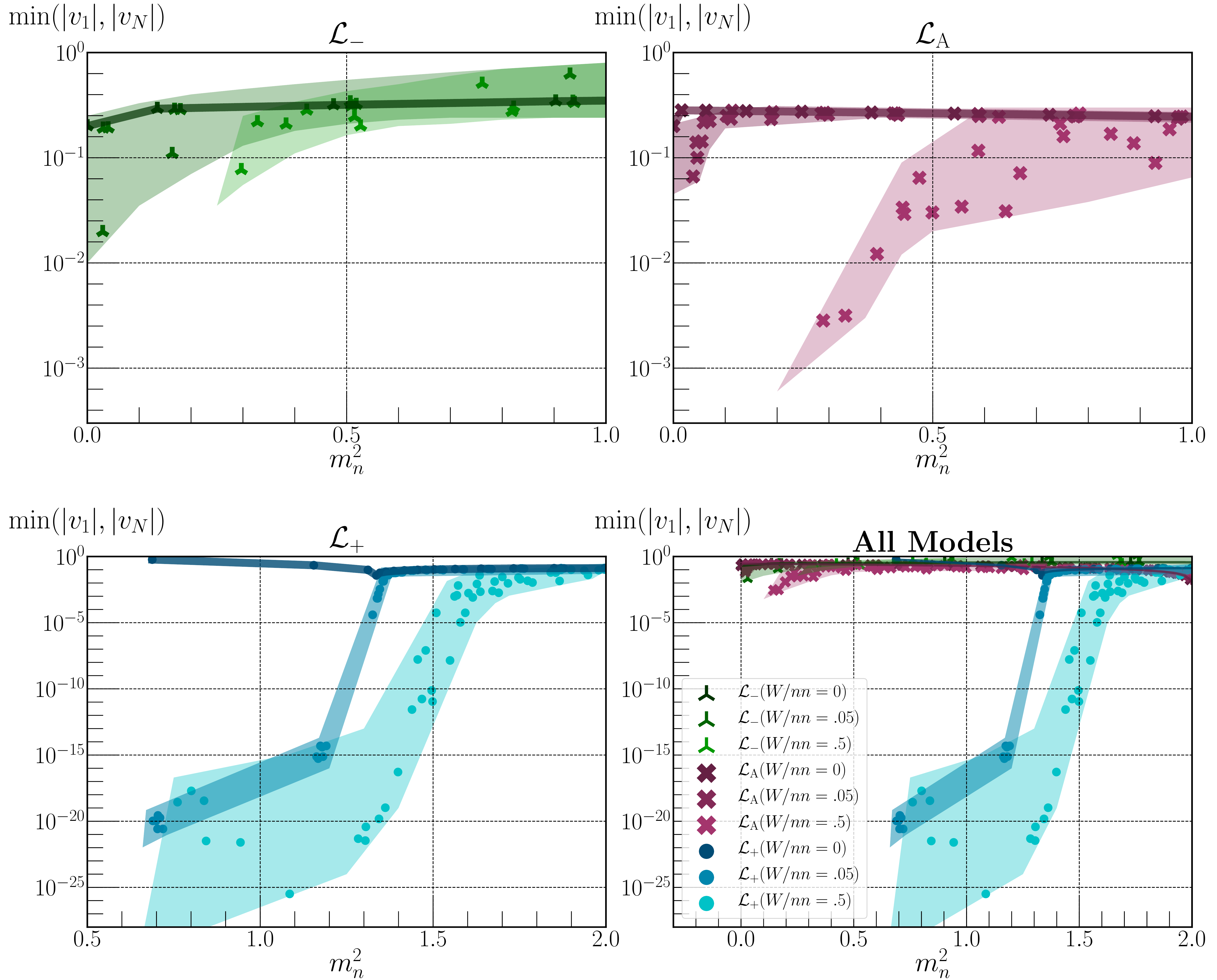}
    \caption{Minimal end-site overlaps as a function of the mass eigenvalues in various toy models. Here, $nn$ stands for 'nearest-neighbor coupling' and is given by $g$ and $g/b$ in $\mathcal{L}_A$ and $\mathcal{L}_{\pm}$, respectively. The first three panels are zoom-ins of each individual model while the last panel combines all of them. Each band indicates the spread of the data from an ensemble of matrices fixing $N=25$. Shaded regions represent typical values from an ensemble of matrices with given $W/nn$, while points denote values for some representative matrices in the ensemble. }
    \label{fig:end-site overlaps}
\end{figure}

So far we focus on the lightest eigenstate and have not yet addressed the localization properties of the other light mass eigenstates. To this end, we present Figure \ref{fig:end-site overlaps}, which plots the minimal end-site overlaps, $\min(|v_1|,|v_N|)$, as a function of mass for the various ensembles. For this plot, we have normalized the couplings $g$'s in the various models to establish a similar bandwidth, while adjusting $W$ to ensure that it is scaled proportionally to the nearest neighbor couplings ($g$ and $g/b$ in the $\mathcal{L}_{\text{A}}$ and $\mathcal{L}_\pm$ models, respectively) in order to provide a fair comparison across the models. Having a small value of $\min(|v_1|,|v_N|)$ indicates that the state is well-localized. One could see that indeed for all the models, the lightest eigenstate is more localized than the other states.

\subsubsection{Semi-analytical Understanding}
\label{sec:analytical considerations}

While $\mathcal{L}_+$ is just a toy model, it has very compelling numerical properties in the limit $W \ll g$, which necessitate a more complete analytical treatment. This is important because as we will show in the next section, while $\mathcal{L}_+$ itself doesn't have a UV completion, UV theories tend to interpolate between $\mathcal{L}_+$ and $\mathcal{L}_-$. Understanding $\mathcal{L}_+$ will be useful to understand the results of more complicated Lagrangians. 

~

\noindent \textit{Background on Bisymmetric Matrices.} Consider $N$ real scalars $\pi_i$ with $i=1,\cdots, N$. In the regime, $W \ll g$, we can write the mass matrix of the $\pi$ fields with $\mathcal{L}_+$ as:
\begin{equation}
    M_\pi = P +  C,
\end{equation}
where $P$ and $C$ correspond to the random diagonal perturbations and contributions from $(\pi_i+\pi_j)^2$ terms respectively: 
\beq
\begin{split}
    P_{i,j} &= \epsilon_i \delta_{i,j} \\
    C_{i,j} &=g\left[b^{-|i-j|} + \frac{3-b - b^{1-i} - b^{i-N}}{b-1} \delta_{i,j}\right],
    \label{eq:defofPC}
\end{split}
\eeq
where the second term in $C$ arises from summing over geometric series, $\sum_{j={i+1}}^{N} b^{-(j-i)} + \sum_{j=1}^{i-1} b^{-(i-j)} -1$. 
The matrix $C$ is almost a Toeplitz matrix with a translational invariance, except for the correlated but not necessarily equal diagonal entries, which break the invariance.

The most important feature of $C$ is that it is bisymmetric. It is both symmetric and persymmetric, which means that $C^F = C$ where $F$ denotes the flip transpose -- a transpose about the diagonal extending from bottom left to top right of the matrix. An even-dimensional bisymmetric matrix could be decomposed into blocks as follows:
\begin{equation}
    C = \begin{pmatrix} A^F & B^T \\ B & A \end{pmatrix},
    \label{def of A, BR}
\end{equation}
where $A^T = A$ and $B^F = B$. The concrete forms of $A$ and $B$ are given in Appendix~\ref{sec:near commutativity}. There is a similar decomposition for odd-dimensional bisymmetric matrices. For simplicity, we will focus on the even-dimensional case in the following discussions and similar arguments could be found for the odd-dimensional case. 

It is useful to introduce an $2N \times 2N$ exchange matrix $J$, 
\beq
J = \begin{pmatrix} 0 & 0 & \cdots & 0 & 1 \\
0 & 0 & \cdots & 1 &0 \\
\vdots & \vdots &\ddots &\vdots & \vdots \\
0 & 1 & \cdots & 0 & 0 \\
1&0 & \cdots & 0 & 0  \end{pmatrix}, 
\eeq
or equivalently, $J_{i,j} = \delta_{i,N+1-j}$. The effect of multiplying $J$ to be left of an arbitrary matrix, $X$, is to flip its rows from top to bottom while the effect of multiplying it from the right is to flip $X$'s columns. It could also be easily verified that $X^F = (JXJ)^T = JX^TJ$. It follows that all bisymmetric matrices commute with $J$. Thus $C$ and $J$ could be simultaneously diagonalized. We call an eigenvector $x$ even if $Jx = +x$ and odd if $Jx = -x$. In other words, even eigenvectors have $x_{j} = x_{N+1-j}$ and are center-symmetric while odd eigenvectors are characterized by $x_{j} = -x_{N+1-j}$ and are center-antisymmetric. 

It is also known that eigenvectors of the bisymmetric matrix $C$ take the form
\begin{equation}
    \begin{pmatrix} Jv \\ v \end{pmatrix} \hspace{10pt},\hspace{10pt} \begin{pmatrix} -Jv' \\ v' \end{pmatrix},
    \label{eq:eigenvectors}
\end{equation}
where $v$ is an eigenvector of $A+BJ$ and $v'$ is an eigenvector of $A-BJ$. These forms are consistent with what we discuss above - the first type of eigenvector is even while the second type is odd. Clearly eigenvectors of $C$ cannot be localized (yet they could be bi-localized as we will show). Given that they are either symmetric or anti-symmetric about the center, they cannot be fit by a single exponential envelope of any size. The eigenvalues of $C$ are given by the union of the eigenvalues of $A+BJ$ and $A-BJ$. Formal proofs of these statements are given in refs.~\cite{brookes2005matrix, reid1997classroom}, which also discuss similar cases of odd dimensional matrices.

~

\noindent \textit{Two Special Properties.} The discussions above only rely on the bisymmetric nature of the $C$ matrix. They hold for all the models such as the $(\pi_i \pm \pi_j)^2$ model. Yet the $C$ matrix representing the $(\pi_i + \pi_j)^2$ model also has some special features, which are absent in the other models. We first summarize these features, which are confirmed numerically: 
\begin{itemize}
\item The two lightest eigenvectors peak at both end sites. 
\item The two lowest eigenvalues of $C$ matrix are almost degenerate, with the splitting decreasing exponentially with $N$: $\log(\lambda_2 - \lambda_1) \sim - N$. 
\end{itemize} 

\begin{figure}[h]
    \centering
    \includegraphics[width=\textwidth]{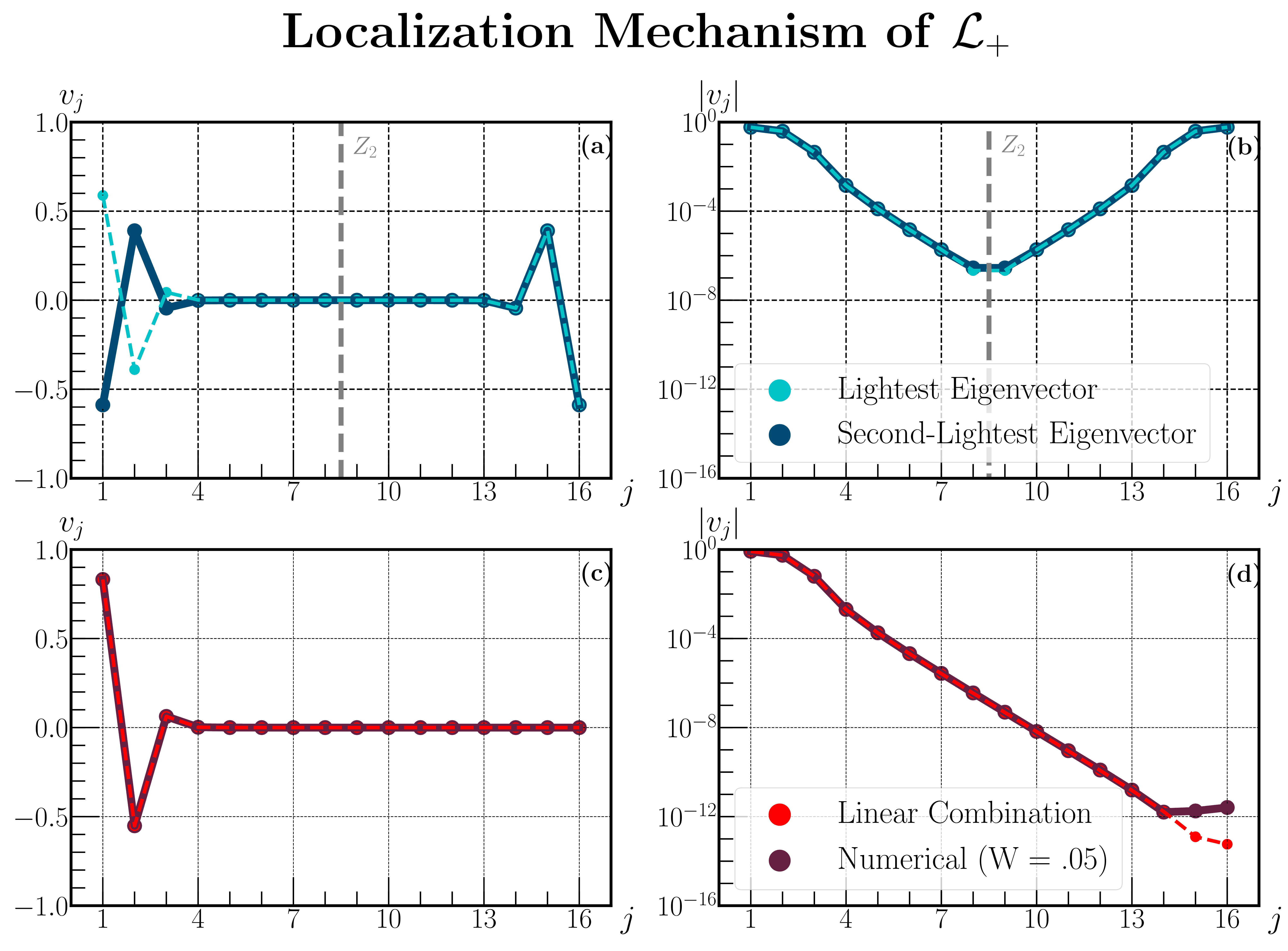}
    \caption{Localization mechanism in the ${\mathcal L}_+$ model. Before adding diagonal randomness, components of the two lightest eigenstates (top left) and their absolute values (top right). After adding randomness, components of new lightest eigenstates compared with the linear combination of two lightest eigenstates from the above panels (bottom left) and their absolute values (bottom right).}
    \label{fig:localization mechanism}
\end{figure}

Below we provide some semi-analytical understandings of the two features. These features arise partly due to the near commutativity of the submatrices $A$ and $BJ$. More specifically, $A$ and $BJ$ commute in the infinite $N$ limit. At a finite $N$, there exists a unitary matrix, $U$, that could almost diagonalize both $A$ and $BJ$, with the average of the off-diagonal components of $U^\dagger A U$ and $U^\dagger (BJ) U$ scaling as a negative power of $N$. We provide the mathematical proof in Appendix~\ref{sec:near commutativity}. Thus $A$ and $BJ$ share similar eigenvectors with small differences at large but finite $N$. Consequently, $v$ and $v^\prime$, the lightest eigenvectors of $A+BJ$ and $A-BJ$ respectively are almost identical as well. Indeed numerically we confirm that at least for the two lightest eigenstates, $v \approx v^\prime$. In addition, numerical studies show that $v$ and $v^\prime$ peak at the $N$-th site (or equivalently, they are dominantly supported by $\pi_N$) with an approximate exponential envelope.\footnote{We don't have a simple analytical argument for that.} Given Eq.~\eqref{eq:eigenvectors}, the two lightest eigenstates of the matrix $C$ that encodes $(\pi_i + \pi_j)^2$ are approximately $v_0 = (Jv, v)^T$ and $v_1 \approx (-Jv, v)^T$, both of which peak at both ends. This is shown in the top panels of Figure \ref{fig:localization mechanism}. 

\begin{figure}[h]
    \centering
    \includegraphics[width = .6\textwidth]{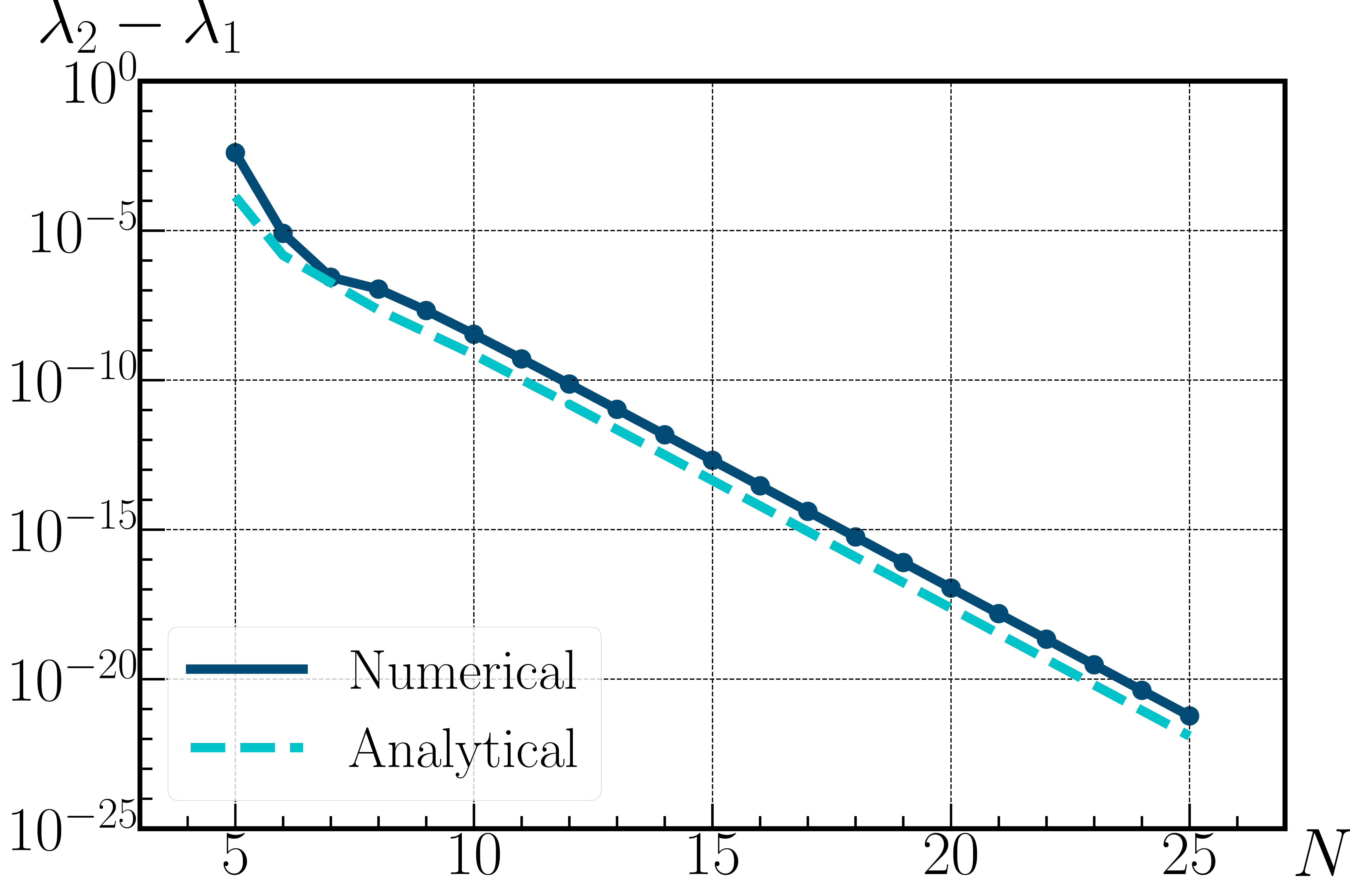}
    \caption{Differences between the two smallest eigenvalues for the model with Lagrangian $\mathcal{L}_+$, fixing $g=1$, $W=0.05$ and $b=2$. Solid blue: exact numerical results. Dashed blue: approximation $2v^T \cdot   (BJ)  \cdot v$. }
    \label{fig:near degeneracy}
\end{figure}

The corresponding two smallest eigenvalues, $\lambda_1, \lambda_2$, are approximately $v^T \cdot \left(A  \pm (BJ) \right) \cdot v$ (assuming $v$'s are normalized) so that the difference between them is  approximately $2 v^T \cdot (BJ) \cdot v$. Indeed this estimate agrees well with the numerical results, as shown in Fig.~\ref{fig:near degeneracy}. The mass splitting exponentially decreases with increasing $N$. 
Yet there is one subtlety in evaluating $2 v^T \cdot (BJ) \cdot v$. One has to use the numerical values of $v$ instead of using the approximate exponential envelop $v \sim e^{(i-N)/L}$. The reason is that since the mass splitting is tiny, any small deviations between the exact values of $v$ and the exponential fits could modify its value.

~

\noindent \textit{Localization Mechanism.} The special properties of $C$ discussed above tell us how localization emerges in the $(\pi_i + \pi_j)^2$ model in the small diagonal disorder limit. When the diagonal perturbations, $\epsilon_i \pi_i^2$ are absent, the two lightest eigenstates, $v_0=(Jv, v)^T$ and $v_1=(- Jv, v)^T$, form a two-dimensional nearly degenerate sub-space with wavefunctions peaking at both end sites. After adding the diagonal perturbations, the mass matrix of $\pi$'s in the subspace becomes 
\begin{equation}
     \begin{pmatrix} \lambda_1 & x \\ x & \lambda_2 \end{pmatrix},  \quad {\rm where} \;  \lambda_1 \approx \lambda_2, \; x = v_0^T \cdot P \cdot v_1 \approx  \frac{1}{2} \left(\epsilon_N - \epsilon_1 \right). 
\end{equation}
Diagonalizing the subspace, one finds that the two new lightest eigenstates are given by $v_0 \pm v_1 = (0,v)^T$ and $(Jv,0)^T$ to the leading order in the perturbation theory. Since $v$ only peaks at one end site, the two lightest eigenstates are exponentially localized around one end site once the perturbation is turned on. This is shown in the bottom panels of Figure \ref{fig:localization mechanism}.

Note that localization emerges in the models with $\mathcal{L}_{-}$ or $\mathcal{L}_{\rm A}$ in a quite different way. For both models, there is no approximate two-fold degeneracy between the two lightest eigenstates. In fact, the lightest eigenstate is massless before adding the diagonal perturbations represented by the matrix $P$. It could be understood most easily from the point of view of UV completions. In both UV completions of the two models, there remains a $U(1)$ symmetry which is only spontaneously broken, resulting in a massless goldstone mode with a flat wavefunction. Turning on a non-zero $P$ leads to the localization of the lightest state. Yet the localization length shrinks slowly with increasing the magnitude of the perturbation $W$. This is in sharp contrast with the $\mathcal{L}_{+}$ model. No matter how small the perturbation is, one gets exponential localization with a small localization length immediately after the perturbation is included because the lightest states are already bi-localized.

\subsection{UV Completions}
 \label{sec:UV completions}
 
The real scalars $\pi$'s could be identified as pseudo Nambu-Goldstone bosons, arising from symmetry breaking of a model with a set of complex scalars $\Phi$'s. The exponentially suppressed mass mixing between non-adjacent $\pi$'s could be generated at loop levels from interactions between adjacent scalars. Let's illustrate the idea with the following Lagrangian of $N$ complex scalars,
\begin{equation}
    \mathcal{L}_1 = \sum_{j=1}^N |\partial_\mu \Phi_j|^2 - V(\Phi_j^\dagger \Phi_j) -\frac{1}{4} \left( \sum_{j=1}^N\epsilon_j \Phi_j^2 + \sum_{j=1}^{N-1} \lambda_j\Phi_{j}^2\Phi_{j+1}^{\dagger 2} + \text{h.c.}\right). 
\label{eq:L1}
\end{equation}
Similar to the UV completion for the Anderson-inspired model in Eq.~\eqref{eqn: C+S UV Completion}, each scalar is charged under an independent $U(1)$ and the entire system transforms under $U(1)^N$. This symmetry is spontaneously broken by the symmetry preserving potential term $V(\Phi_j^\dagger \Phi_j)$ and each scalar obtains a vacuum expectation value (vev). For simplicity, we assume all the vevs are the same, $\langle \Phi_j \rangle = f$. Using non-linear parametrization, we write $\Phi_j = \left(f + \phi_j \right) e^{i \pi_j/ (\sqrt{2} f)}$. 

\begin{figure}[h]
\begin{center}
\includegraphics[width=.6\textwidth]{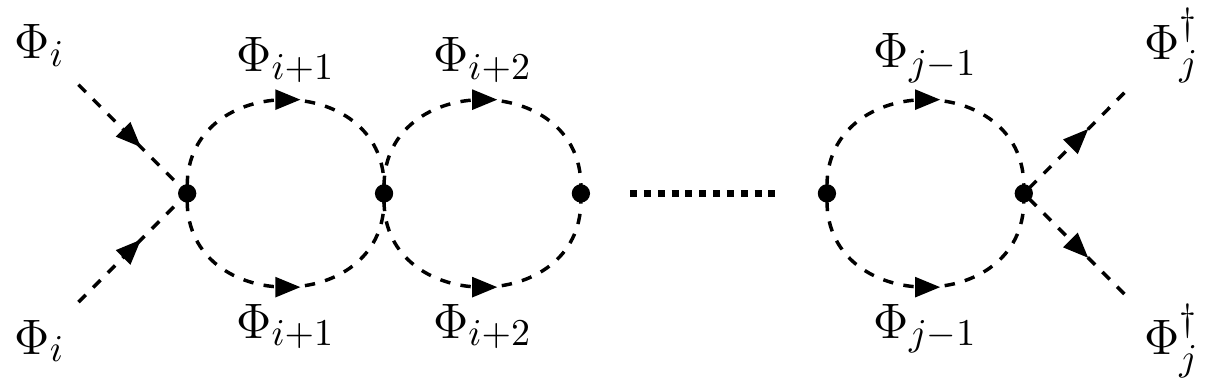}
\caption{An example of the diagrams that generate quartic interaction between two non-adjacent scalars at loop level, $\Phi_i^2 \Phi_j^{\dagger 2}$. }
\label{fig:fydiagram}
\end{center}
\end{figure}

Now we inspect the two explicit symmetry breaking terms and the potential of the pseudo Nambu-Goldstones they generate. We assume $\lambda_j \ll 1$ and $\epsilon_j \ll f^2$ so that these two terms could be treated as perturbations. The $\epsilon_j \Phi_j^2$ terms break the $U(1)^N$ symmetry explicitly and entirely while the quartic interaction between each $\Phi$ and its adjacent scalar breaks $U(1)^N$ explicitly down to a single $U(1)$. Assuming that each scalar carries charge $+1$ under the corresponding $U(1)$, each quartic coupling $\lambda_j$ could be treated as a symmetry breaking spurion, carrying charge $-2$ under $U(1)_j$ and charge $2$ under $U(1)_{j+1}$. At tree level, these two explicit breaking terms generate random diagonal masses, e.g., $\epsilon_j \pi_j^2/2$ and mass mixing between adjacent $\pi$'s, e.g., $\lambda_j f^2 (\pi_j- \pi_{j+1})^2$. At loop level, one could generate quartic interactions between non-adjacent complex scalars such as $\Phi_i^2 \Phi_j^{\dagger 2}$ with $|j-i| >1$. By spurion analysis, the coefficient of $\Phi_i^2 \Phi_j^{\dagger 2}$ has to be proportional to $\lambda_i \lambda_{i+1} \cdots \lambda_j$. One example Feynman diagram is shown in Fig.~\ref{fig:fydiagram}, which leads to an estimate of the coefficient to be 
\beq
\frac{\lambda_i \lambda_{i+1} \cdots \lambda_{j-1}}{(4\pi)^{2(|j-i|-1)}}
\eeq
up to a logarithmic factor. We assume that all $\lambda$'s take the same value for simplicity, $\lambda_j \equiv \lambda$. Integrating out the heavy radial modes (with masses of order $\sim f$) and mapping onto the low-energy EFT containing only $\pi$'s (with masses of order ${\rm max}(\sqrt{\lambda}f, \sqrt{\epsilon}$\,) ), we obtain $\mathcal{L}_-$ in Eq.~\eqref{eq:non-Local Lagrangian1} and identify
\beq
g \sim (4\pi)^2 f^2, \quad b \sim \frac{(4\pi)^2}{\lambda}. 
\eeq
Note that the identifications above are approximate and could differ at the order one level, due to the logarithms from loops.

One may think to apply the same strategy as above to UV complete $\mathcal{L}_+$. However, using spurion analysis, one would find that it is not possible to construct a UV completion leading to $\mathcal{L}_+$. Instead, one always gets a mixture of $(\pi_i+ \pi_j)^2$ and $(\pi_i-\pi_j)^2$ terms. To see this, consider a different tree-level Lagrangian: 
\beq
    \mathcal{L}_2 \supset - \left( \sum_{j=1}^{N-1} y_j\Phi_{j}^2\Phi_{j+1}^{2} + \text{h.c.} \right),
\eeq
where terms other than the quartic interactions are identical to those in Eq.~\eqref{eq:L1}. In this case, the spurion $y_j$ carries charge $-2$ under $U(1)_j$ and $-2$ under $U(1)_{j+1}$. A series of quartic interactions between non-adjacent $\Phi$'s are generated at loop order, taking the form
\beq
&&\frac{y_i y_{i+1}^* y_{i+2} \cdots y_{j-1}}{(4\pi)^{2(|j-i|-1)}} \Phi_i^2 \Phi_j^2 + \text{h.c.} \quad \; {\rm when} \; |j-i| = {\rm odd}; \nonumber \\
&&\frac{y_i y_{i+1}^* y_{i+2} \cdots y_{j-1}^*}{(4\pi)^{2(|j-i|-1)}} \Phi_i^2 \Phi_j^{\dagger 2} +\text{h.c.} \quad {\rm when} \; |j-i| = {\rm even}. 
\eeq \label{L2 operators}
Using the non-linear parametrization of $\Phi$ and setting all $y_i$'s to be real and equal to $y$ to preserve CP, one could find this leads to a low-energy EFT with the Lagrangian
\begin{equation}
    \frac{1}{2} \sum_{i=1}^N (\partial_\mu \pi_{i})^2 - \frac{1}{2} \sum_{j=1}^N \epsilon_j \pi_j^2 - \frac{1}{2} \sum_{i=1}^{N-1} \sum_{j=i+1}^{N} \frac{g}{b^{j-i}} (\pi_i -(-1)^{j-i} \pi_j)^2,
\end{equation}
which could be reduced to $\mathcal{L}_-$ by a field redefinition $\pi_{\text{even}} \mapsto -\pi_{\text{even}}$. 

\begin{figure}[h]
    \centering
    \includegraphics[width = 5in]{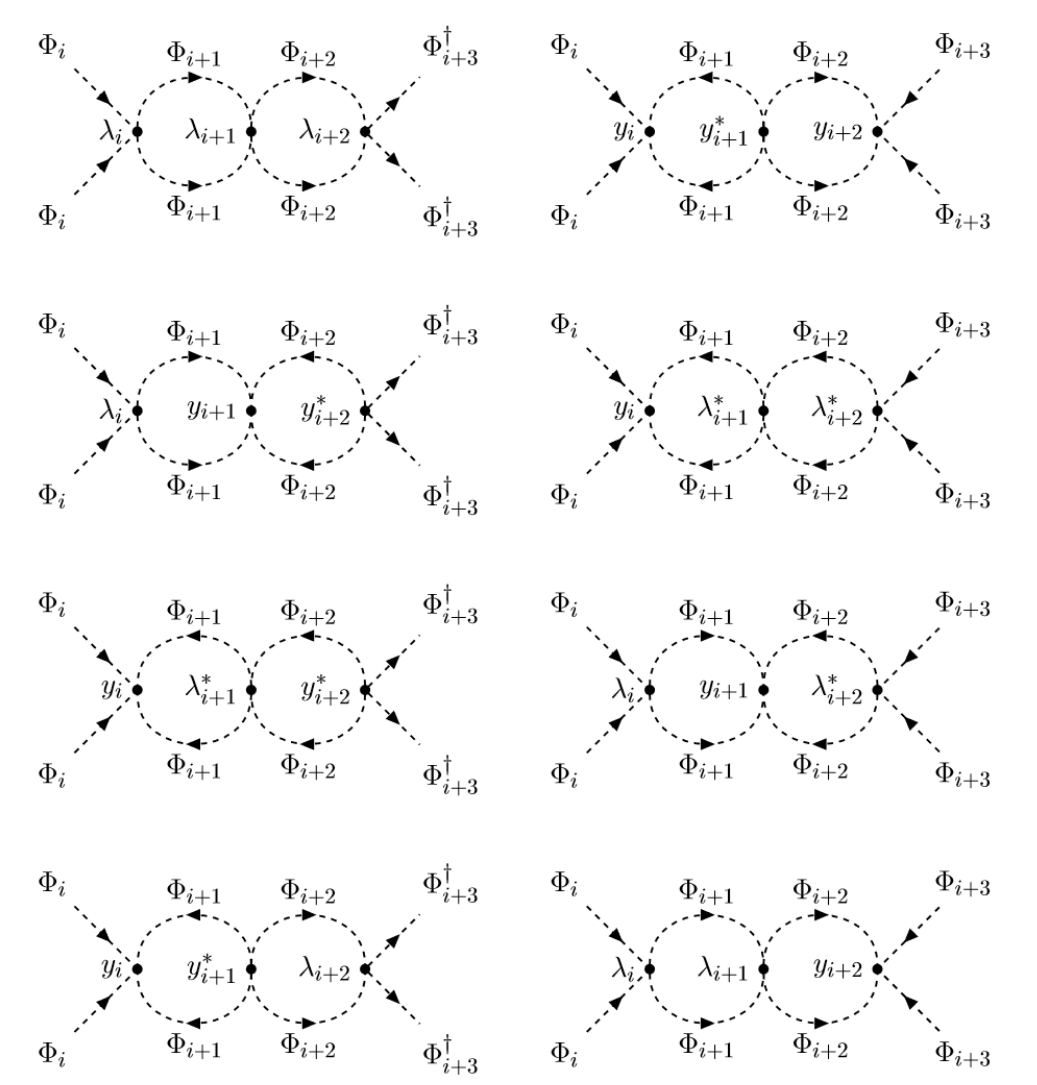}
    \caption{The 8 Feynman diagrams contributing to the lowest order mass-mixing terms between $\Phi_i$ and $\Phi_{i+3}$.}
    \label{fig:8FeynmanDiagrams}
\end{figure}

One could also consider a mixture of quartic interactions containing both $\Phi_{j}^2\Phi_{j+1}^{2}$ and $\Phi_{j}^2\Phi_{j+1}^{\dagger2}$. This leads to a low-energy EFT which contains a linear combination of the potential terms in $\mathcal{L}_-$ and $\mathcal{L}_+$. Notably, while each individual quartic interaction explicitly breaks the $U(1)^N$ symmetry down to $U(1)$, considering both sets of interactions together breaks the $U(1)^N$ completely. To derive the form of this Lagrangian, we could again perform a spurion analysis. Take the mass-mixing between $\pi_i$ and $\pi_{i+3}$ as an example. There are eight Feynman diagrams contributing, as depicted in Figure \ref{fig:8FeynmanDiagrams}. They generate
\begin{equation}
\begin{split}
    &\frac{1}{(4\pi)^4} \big(\lambda_i \lambda_{i+1}\lambda_{i+2} + \lambda_i y_{i+1} y_{i+2}^* + y_i \lambda_{i+1}^* y_{i+2}^* + y_i y_{i+1}^* \lambda_{i+2} \big)\Phi_i^2\Phi_{i+3}^{\dagger2}+ \\
    & \frac{1}{(4\pi)^4} \big(y_i y_{i+1}^* y_{i+2} + y_i \lambda_{i+1}^* \lambda_{i+2}^* + \lambda_i y_{i+1} \lambda_{i+2}^* + \lambda_i \lambda_{i+1} y_{i+2} \big)\Phi_i^2\Phi_{i+3}^{2} + \text{h.c}
\end{split}
\end{equation}
up to logarithms.
Now, we take $\lambda_j \equiv \lambda$ and $y_j \equiv y$ with $\lambda,y$ real to simplify the expression. The above equation then becomes
\begin{equation}
    \frac{1}{(4\pi)^4} \bigg[ \big(\lambda^3 + 3 \lambda y^2\big)\Phi_i^2\Phi_{i+3}^{\dagger2} + \big(3\lambda^2 y + y^3\big)\Phi_i^2\Phi_{i+3}^{2}\bigg] + \text{h.c.}
\end{equation}
Note that the prefactor of $\lambda y^2$ ($\lambda^2 y$), which is three in this example, could be understood as a combinatoric ${3 \choose 1}$: we are choosing exactly one of the three vertices to be a $\lambda$ ($y$). In general, the coefficient of the mass mixing term takes the form $\lambda^a y^b$ with $a+b$ equal to the separation of the two scalar fields. The prefactors could be obtained using a combinatorial argument. First go back to the original model with the most general couplings. In the spurion analysis, $y_i y_j^*$ carries charge (-2, 2) under $U(1)_i \times U(1)_{j+1}$. Equivalently, a diagram going through $y_i$ and $y_j^*$ vertices connect $\Phi_i^2$ and $\Phi_{j+1}^{\dagger 2}$. In general, diagrams containing even pairs of $y_i$ and $y_j^*$ vertices connect a scalar field and the complex conjugate of another scalar field. On the other hand, diagrams containing odd numbers of $y_i$ and $y_j^*$ vertices connect a scalar field with another scalar field or the complex conjugate with another complex conjugate. When we set all $y$'s ($\lambda$'s) to be equal and real, this just reduces to the statement that if the coefficient contains even numbers of $y$'s ($b$ is even), they are associated with $\Phi_i^2 \Phi_j^{\dagger 2}$ while they contain odd numbers of $y$'s ($b$ is odd), they are associated with $\Phi_i^2 \Phi_j^{2}$. The prefactors could be obtained by counting the number of ways to select $b$ vertices out of a total of $a+b$ vertices to be $y$'s: ${a+b \choose b}$.

This analysis gives rise to the following quartic operators connecting $\Phi_i$ and $\Phi_j$ (or the conjugates):
\begin{equation}
    \frac{1}{(4\pi)^{2(|j-i|-1)}}\bigg[ \sum_{m \text{~even}} \begin{pmatrix} |j-i| \\ m \end{pmatrix} y^m \lambda^{|j-i|-m} \Phi_i^2\Phi_{j}^{\dagger2} + \sum_{m \text{~odd}} \begin{pmatrix} |j-i| \\ m \end{pmatrix} y^m \lambda^{|j-i|-m} \Phi_i^2\Phi_{j}^{2}\bigg] + \text{h.c.}
\end{equation}
up to logarithmic factors we ignore. The reason we could ignore logarithmic factors is that varying the coefficients by order one amount does not change the localization property, as we will show in the following section. 
$\Phi_i^2\Phi_j^2$ lead to $(\pi_i + \pi_j)^2$ while $\Phi_i^2\Phi_j^{\dagger 2}$ lead to $(\pi_i - \pi_j)^2$ in the low-energy EFT. Combining and summing over terms lead to the following Lagrangian of $\pi$'s: 
\begin{equation}
    \mathcal{L}_{\rm mixed} = \frac{1}{2} \sum_{i=1}^N (\partial_\mu \pi_{i})^2 - \frac{1}{2} \sum_{j=1}^N \epsilon_j \pi_j^2 - \frac{g}{2} \sum_{i=1}^{N-1} \sum_{j=i+1}^{N} \Big( \frac{1}{c^{|i-j|}}(\pi_i^2 + \pi_j^2)  - \frac{2}{d^{|i-j|}}\pi_i \pi_j \Big),
\end{equation}\label{L EFT}
where we could identify: 
\begin{equation}
    g = (4 \pi)^2 f^2, \quad c = \frac{(4 \pi)^2}{\lambda + y}, \quad d = \frac{(4 \pi)^2}{\lambda - y}.
    \label{eq:mapping}
\end{equation}

It is possible to generalize the discussions above in two ways. The first is to consider the case where we allow the couplings $y$'s ($\lambda$'s) to take different values. Indeed, one could conceivably let $\epsilon_i,y_i,\lambda_i$ all be random parameters pulled from various distributions. We shall treat this setting in the following section. Another way to generalize our discussion is to consider a UV theory will all possible relevant and marginal operators that couple adjacent scalars in the theory space. For example, one could consider a UV theory that also contains operators like $\Phi_i^2 \Phi_i^\dagger \Phi_{i+1}$ and $\Phi_i^2 \Phi_{i+1}$. Interestingly, allowing for the zoo of operators gives qualitatively similar results when we consider (non-local) mass-mixing terms to their lowest order in couplings because for any set of operators that allow us to traverse through the theory space via loops, the loop structures are similar and getting closed form answers is reduced to combinatorics. Because the models appear to be qualitatively similar, we forgo discussing this more complicated example.

\subsection{Localization of $ \mathcal{L}_{\rm mixed}$}
\label{sec:Lmixed}

 \begin{figure}[h]
     \centering
     \includegraphics[width = .6\textwidth]{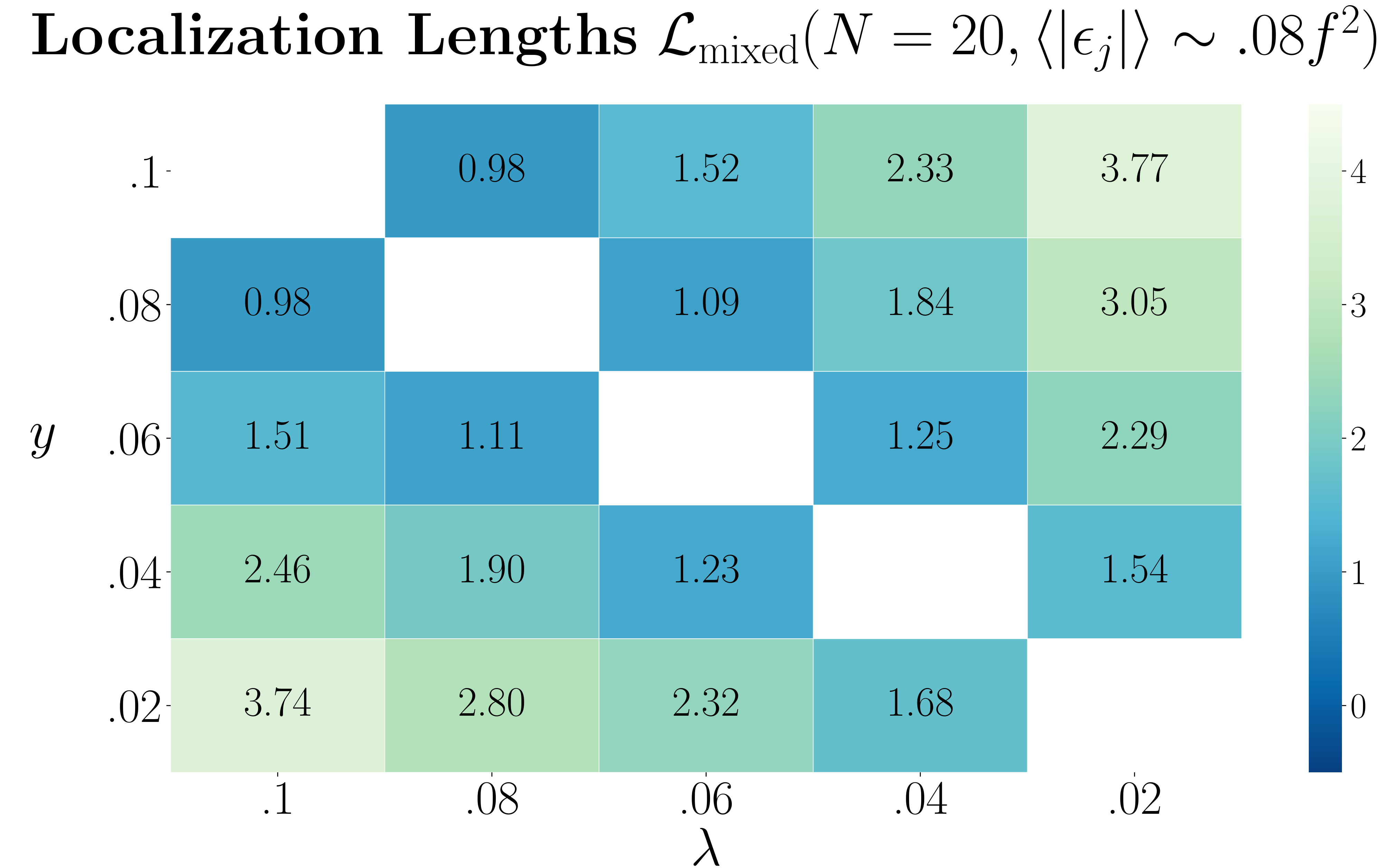}
     \caption{Median localization length heat map (based on ensembles of 100 matrices for each set of $\lambda$ and $y$) for $\mathcal{L}_{\text{mixed}}$ with $\langle \epsilon_j \rangle \sim 0.08 f^2$. Note that we have \textit{not} taken the $\log_{10}$ of the localization lengths in this figure as we have done with the other heat maps. Accordingly, one should be cautious to compare this heat map to the others presented earlier.}
     \label{fig:EFT Heatmap}
 \end{figure}

In this section, we will show that the $\mathcal{L}_{\text{mixed}}$ model constructed at the end of last section allows us to achieve $\mathcal{O}(1)$ localization lengths even when randomness is arbitrarily small. The mechanism that guarantees localization in the small randomness region is identical to the mechanism for the $\mathcal{L}_+$ model discussed in section~\ref{sec:analytical considerations}.

 We first note that UV Lagrangian described in the last section that leads to $\mathcal{L}_{\text{mixed}}$ has four free parameters: $W,f,\lambda,$ and $y$. $\mathcal{L}_{\text{mixed}}$ could be parametrized by $W,g,c,d$, which are functions of the UV parameters, as demonstrated in Eq.~\eqref{eq:mapping}. For the rest of the section, we will use the set of UV parameters. To have the explicit symmetry breaking terms controlled by $W, \lambda$ and $y$ as perturbations and all quartic couplings relevant for the discussions, we need to impose
 \begin{equation}
     \begin{split}
       W \ll f^2, \\
       \lambda \ll 1,\\
         y \ll 1, \\
         \lambda \sim y.
     \end{split}
 \end{equation}
The small parameters are technically natural in the sense that in the limit they are zero, the Lagrangian preserves a $U(1)^N$ symmetry. 
For the rest of the section, we take $W/g =10^{-3}$, which corresponds to $\langle \epsilon_j \rangle /f^2 \approx 0.08$, and vary $\lambda$ and $y$. 
 
The heat map of median localization length for this model is presented in Figure \ref{fig:EFT Heatmap}. Note that while all other heat maps in this article present $\log_{10} (L)$, we simply present the median localization lengths in Figure \ref{fig:EFT Heatmap}. Remarkably, all these median localization lengths are all $\mathcal{O}(1)$ when $\lambda$ and $y$ are of order of $10^{-2} - 10^{-1}$. Note that the localization lengths should remain the same under the exchange $(\lambda,y) \mapsto (y,\lambda).$ This is because this transformation changes $\mathcal{L}_{\text{mixed}}$ via $(c,d) \mapsto (c,-d)$, which is the same as $\mathcal{L}_{\text{mixed}}$ under the field redefinition $\pi_{\text{even}} \mapsto - \pi_{\text{even}}.$ The small differences in the localization length under the exchange, shown in Fig.~\ref{fig:EFT Heatmap}, is due to statistical fluctuation of our ensemble with a limited number of matrices.

  \begin{figure}[h]
     \centering
     \includegraphics[width = \textwidth]{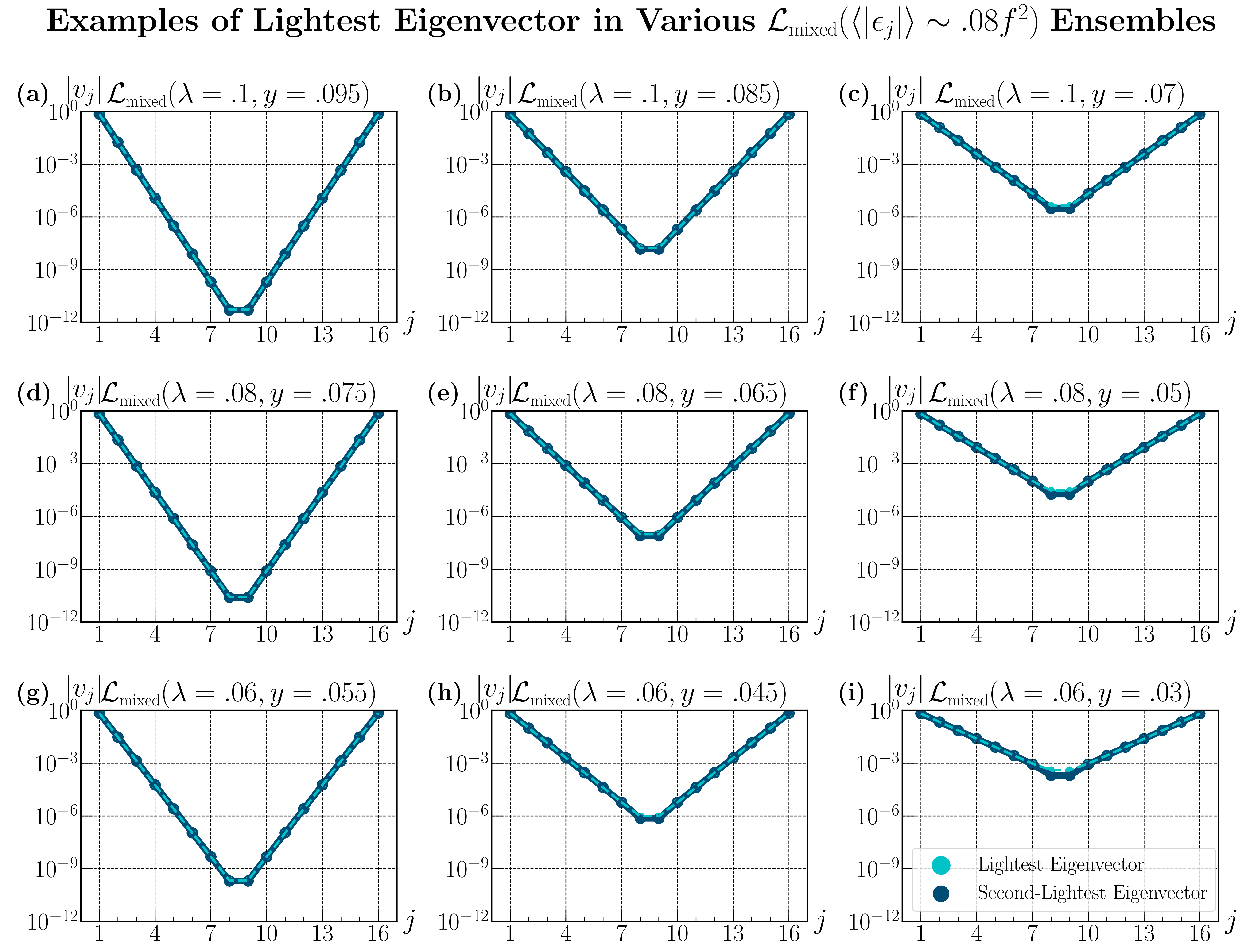}
     \caption{The absolute values of the components of the lightest two eigenstates in various $\mathcal{L}_{\text{mixed}}$ models.}
     \label{fig:EFT Lightest Eigenstates}
 \end{figure}
 
  \begin{figure}[h]
     \centering
     \includegraphics[width =0.8 \textwidth]{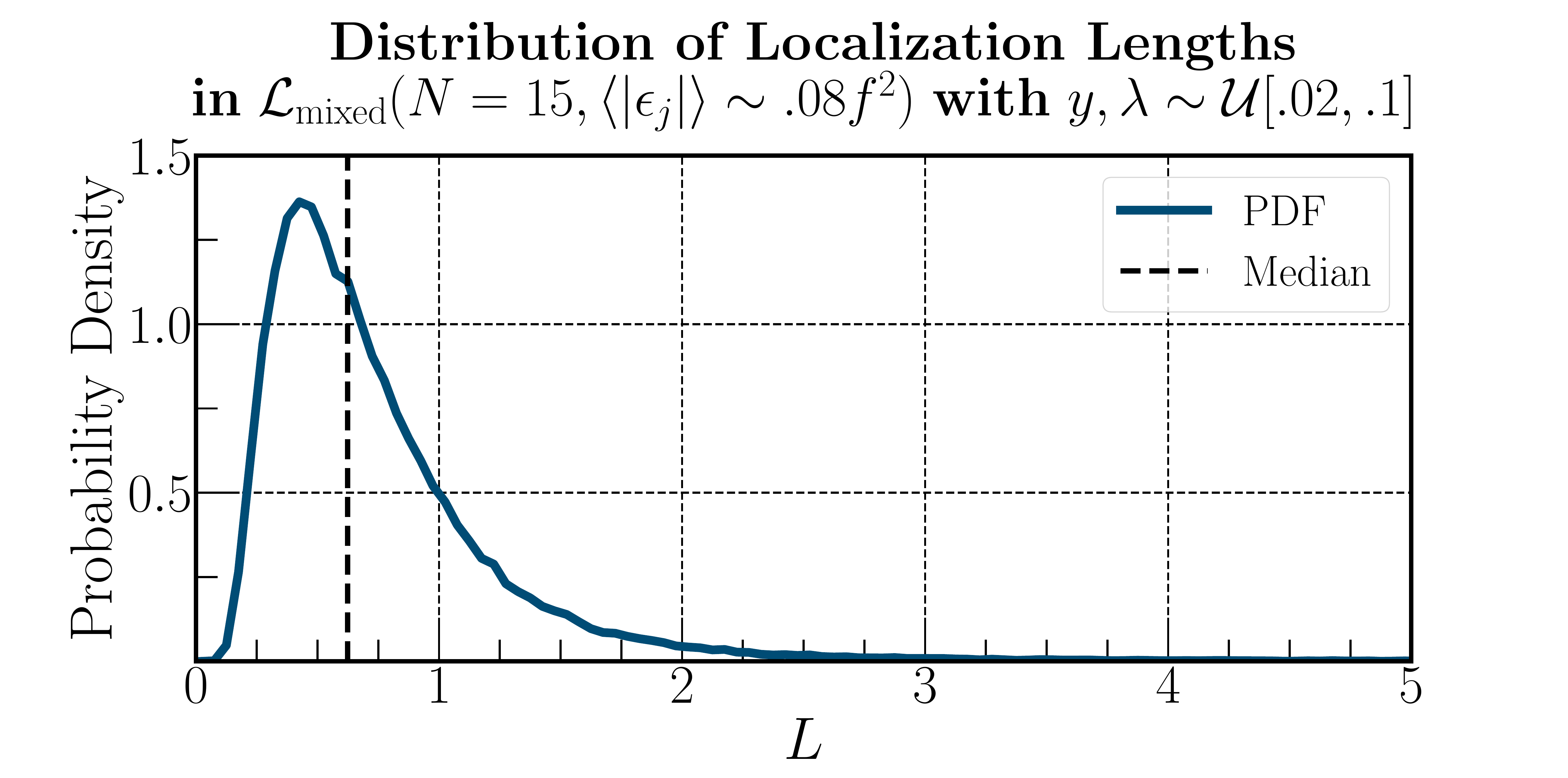}\quad
     \includegraphics[width =0.8  \textwidth]{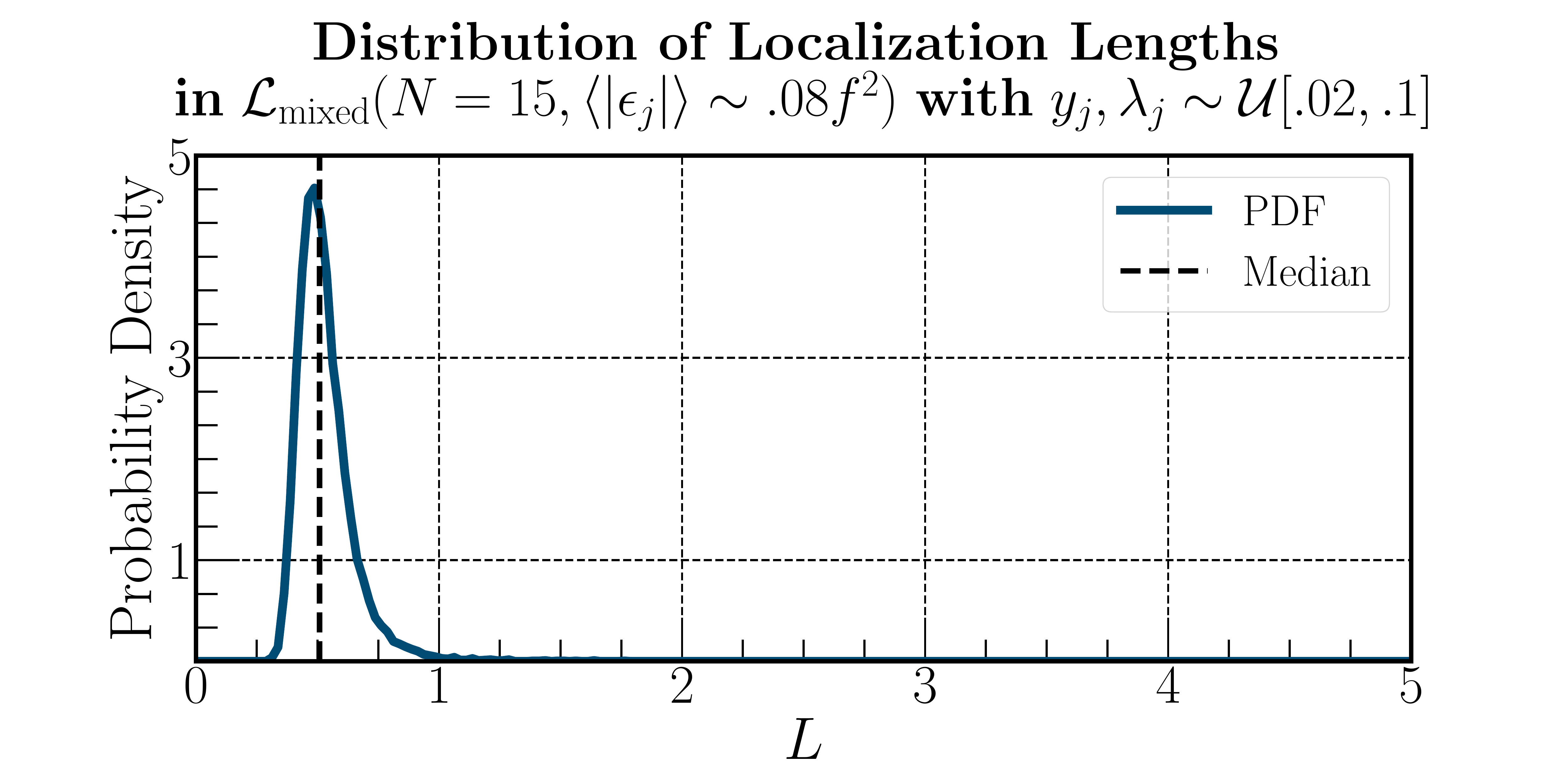}
     \caption{Top: distribution of localization lengths for $\mathcal{L}_{\text{mixed}}$ with $\lambda_j \equiv \lambda$ and $y_j \equiv y$ and $\lambda,y$ taken independently from a uniform distribution $\mathcal{U}[.02,.1]$. Bottom: distribution of localization lengths for $\mathcal{L}_{\text{mixed}}$ with $\lambda_j \not\equiv \lambda$ and $y_j \not\equiv y$ and $\lambda_j,y_j$ all taken independently from $\mathcal{U}[.02,.1]$. Distributions are computed using 100,000 trials and 10,000 trials respectively.}
     \label{fig:EFTDistributionsConstant}
 \end{figure}
 
 Why do these $\mathcal{O}(1)$ localization lengths appear even when $W/g \ll 1$? It turns out that the localization mechanism for $\mathcal{L}_{\text{mixed}}$ is identical to the localization mechanism in $\mathcal{L}_+$ studied in section~\ref{sec:analytical considerations}. Indeed, we see that the two lightest states of $\mathcal{L}_{\text{mixed}}$ with $W = 0$ are nearly parity-conjugates of one another with maximal support on the first and the last components. This is depicted in Figure \ref{fig:EFT Lightest Eigenstates}, which plots the two lightest eigenstates for a random selection of this class of models. Furthermore, the masses of these states are nearly identical with a tiny difference for many technically natural values of $\lambda$ and $y$. It follows that the conditions described in section \ref{sec:analytical considerations} are met and that for arbitrarily small values of $W$, the lightest eigenstate will be exponentially localized with support around one particular end-site.

 What happens when we loosen our restrictions on $\lambda$ and $y$ by allowing them to assume random values? We first consider the case where $\lambda$ and $y$ are real parameters (to preserve CP), chosen independently from the uniform distribution $\mathcal{U}[.02,.1].$ We give the probability density function (PDF) of the resulting localization lengths in the top panel of Figure \ref{fig:EFTDistributionsConstant}. Interestingly, we find that the median localization length is less than 1 and that $99.9\%$ of all computed localization lengths are less than $10$. Next, recall that in the previous section we made the simplification $\lambda_j \equiv \lambda$ and $y_j \equiv y$ in order to simplify the formalism and to allow a low-energy EFT to be written in a closed form. In principal, however, we do not need to do this; indeed, while we cannot write the mass-matrix of the low-energy EFT in a closed form, we can always construct it and simulate its mass matrix numerically. To this end, we let $\lambda_j$ and $y_j$ be real parameters all pulled independently from $\mathcal{U}[.02,.1].$ The relevant PDF is given in the bottom panel of Figure \ref{fig:EFTDistributionsConstant}. One can check that the median localization length is even smaller in this generalization of the model and, again, $99.9\%$ of all localization lengths are less than $10.$ Altogether, it is clear that the strong localization in the low-energy EFT is robust to random perturbations of the UV theory provided couplings in the UV theory remain small.

\section{Conclusion and Outlook}
\label{sec:conclusion}

In this article, we explore a new class of random matrix model with long-range hopping that decays exponentially with site separation. We first consider the associated Hamiltonian model of a single particle on a 1D lattice with $N$ sites and prove the exponential localization of the energy eigenstates, both numerically and analytically. We then mainly focus on the inspired non-local theory space constructions containing a lattice of scalars, $\pi$'s, mixing with each other in mass terms and with Lagrangians, ${\mathcal L}_\pm$ (Eq.~\eqref{eq:non-Local Lagrangian1}). In particular, we are interested in the small diagonal randomness limit, in which the diagonal random masses of the scalars are small and intuitively there is no guarantee of a localized eigenstate.
 While the various theory space models look similar and all exhibit exponential localization analogous to the Hamiltonian model, they have dramatically different properties. The main findings of our paper are, compared to the local theory space model in Ref.~\cite{Craig:2017ppp}, in the small diagonal disorder limit, 
\begin{itemize}
\item The non-local mixing in the ${\mathcal L}_-$ model weakens the localization of the lightest mass eigenstates.
\item The non-local mixing in the ${\mathcal L}_+$ model significantly strengthens the localization of the lightest mass eigenstates, with an ${\cal O}(1)$ localization length. 
\end{itemize}
What happens in the ${\mathcal L}_+$ model is that in the absence of random diagonal perturbations, the two lightest mass eigenstates are nearly degenerate and peak at both end sites. Once the random diagonal masses are turned on, no matter how small they are, both the lightest eigenstates are mixed up and the resulting new eigenstates peak only at one end site. Unfortunately while the ${\mathcal L}_-$ model could be easily UV completed by identifying $\pi$'s as pNGBs and generating non-local interactions at loop levels, the ${\mathcal L}_+$ model does not have a simple perturbative UV completion. However, we could construct models, following a similar strategy to UV complete the ${\mathcal L}_-$ model, which leads to a low energy EFT of $\pi$'s, ${\mathcal L}_{\rm mixed}$, that interpolates between ${\mathcal L}_+$ and ${\mathcal L}_-$. We show that their localization properties are very similar to the ${\mathcal L}_+$ model. 

There are several directions to expand the study of localization in the non-local theory space models.
One important question, which we don't fully address, is the origin of the exponential bi-localization of the ${\mathcal L}_+$ and ${\mathcal L}_{\rm mixed}$ models in the absence of random disorders. 
Ultimately, the goal of studying exponential localization in the theory space models is to develop new mechanisms to generate exponential hierarchies. For example, as discussed in Ref.~\cite{Craig:2017ppp}, one could use the Anderson-inspired local theory space model to explain the flavor hierarchy of the standard model quarks with an exponential profile of a localized scalar. Similarly, we could use the non-local theory space model to explain the CKM matrix with fewer scalars involved. Other possible applications could be found by drawing analogies to those of generalized clockwork~\cite{Giudice:2016yja} and clockwork WIMP/neutrino models~\cite{Hambye:2016qkf, Park_2018, Ibarra_2018, Hong:2019bki}. Since it is not our focus to develop any new applications in this article, we will save this as exercises for interested readers. It is also of great interest to explore whether there are other classes of random matrix models and associated theory space constructions that enjoy different localization properties, like the ${\mathcal L}_+$ model and its variant we present.

\section*{Acknowledgements} 
We thank Matt Reece for reading the manuscript and providing useful comments. AT is supported by Karen T. Romer Undergraduate Teaching and Research Awards (UTRAs) at Brown and JF is supported by the DOE grant DE-SC-0010010 and NASA grant 80NSSC18K1010. JF thanks KITP at UCSB (supported by the National Science Foundation under Grant No. NSF PHY-1748958) at which part of the work was carried out.

\appendix 

\section{Spectrum of Hamiltonian in Eq.~\eqref{eq:nonlocalH2}}
\label{sec: bounded spectra (appendix)}

In this appendix, we discuss the spectrum of Hamiltonian in Eq.~\eqref{eq:nonlocalH2}. The Hamiltonian could be written as 
\begin{equation}
\begin{split}
H_{\rm long-range}& = H_0 + g T, \\
(H_0)_{ j,k} &= \epsilon_j \, \delta_{j,k}, \\
T_{j,k} &= b^{-|j-k|} (1 - \delta_{j,k}), 
 \end{split}
\label{eq:Split correlated matrix}
\end{equation}
The hopping matrix $T$ is a Toeplitz matrix, respecting the translational invariance $T_{j,k} = t_{j-k}$. It is useful to compute the Fourier transform of the Toeplitz matrix elements~\cite{ekstrom2018eigenvalues}
\beq
    f(\theta) = \sum_{j=-\infty}^\infty t_{j}e^{ij\theta},
\eeq
where the function $f$ is usually referred to as the \textit{symbol} of the Toeplitz matrix. For the Toeplitz matrix defined in Eq.~\eqref{eq:Split correlated matrix}, we find that 
\begin{equation}
    f(\theta) =  \frac{(b^2-1)g}{(b^2+1) -2 b \cos \theta}  -g
    \label{eq:ftheta}
\end{equation}

\begin{figure}[h]
    \centering
    \includegraphics[width=0.6\textwidth]{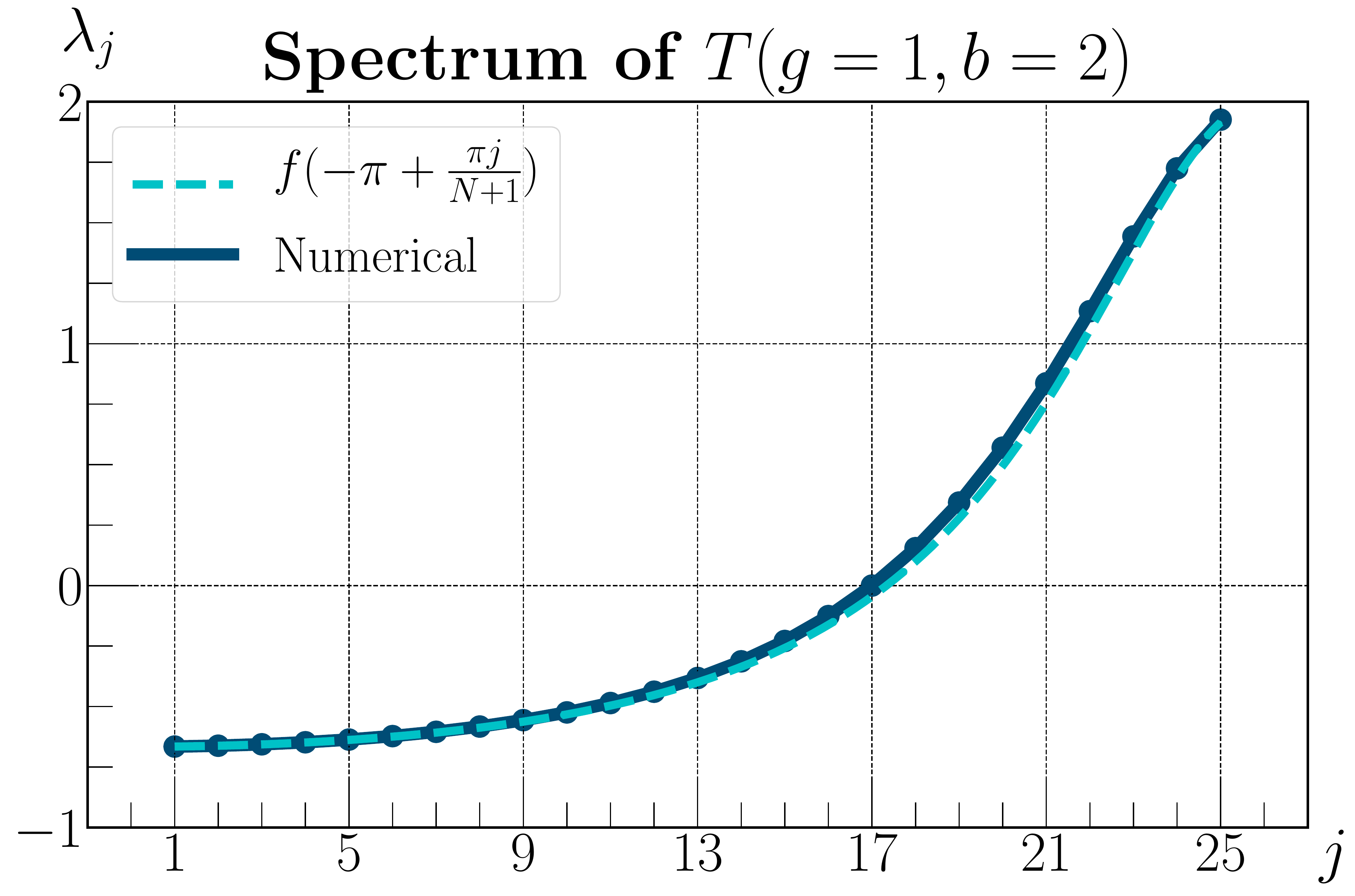}
    \caption{Eigenvalues of the Toeplitz matrix $T$ in Eq.~\eqref{eq:Split correlated matrix}, fixing $N=25$, $g=1$ and $b=2$. Blue dashed line: $f \left(-\pi + \frac{\pi j}{N+1} \right)$; blue solid: numerical results. }
    \label{fig:eigenvalues}
\end{figure}

It is known that the spectrum of a Toeplitz matrix is bounded by the essential infimum and supremum of its symbol~\cite{serra1998extreme}. In our case, since $f$ (Eq.~\eqref{eq:ftheta}) is an even function that is $2\pi$-periodic, we only need to consider $f$ on $[-\pi, 0]$. In the limit $N \rightarrow \infty$, the minimal and maximal eigenvalues converge to $\min f$ and $\max f$ respectively. In addition, for Riemann integrable $f$ such as the one in Eq.~\eqref{eq:ftheta}, the set of eigenvalues $\{\lambda_j\}$ has the same distribution as the set
\begin{equation}
\bigg\{ f \bigg( -\pi + \frac{\pi j}{N+1} \bigg) \bigg\}, \quad j = 1, 2, \cdots, N. 
\end{equation}
Furthermore, it has been proved that the above two sets are almost equal and the difference between corresponding elements is small\cite{ekstrom2018are}. An example is presented in Fig.~\ref{fig:eigenvalues}, where we illustrate that the eigenvalues of $T$ are, indeed, almost identical to the analytical prediction obtained by considering $T$'s symbol.

To compute the bounds on the spectra of the full matrix $H$, one could also use the following inequality:
\beq
    \min (\lambda(H_0))  \leq  \lambda_j(T+H_0) - \lambda_j(T) &\leq & \max (\lambda(H_0)),
\eeq
where $\lambda_j(A)$ denotes the $j$-th largest eigenvalue of matrix $A$ \cite{Horn2012MatrixAnalysis}. 
Altogether, we find 
\beq
    \Delta(H) &\subset & [\min(f) + \min(\lambda(H_0)), \; \max(f) + \max(\lambda(H_0)) ], \nonumber \\
  &= & \left[-\frac{2}{b+1} g, \; \frac{2}{b-1}g + W\right], \nonumber \\
    \label{eq:spectrum bound}
\eeq
which gives Eq.~\eqref{eq:deltaH}. 

\begin{figure}
     \centering
     \includegraphics[width = .6\textwidth]{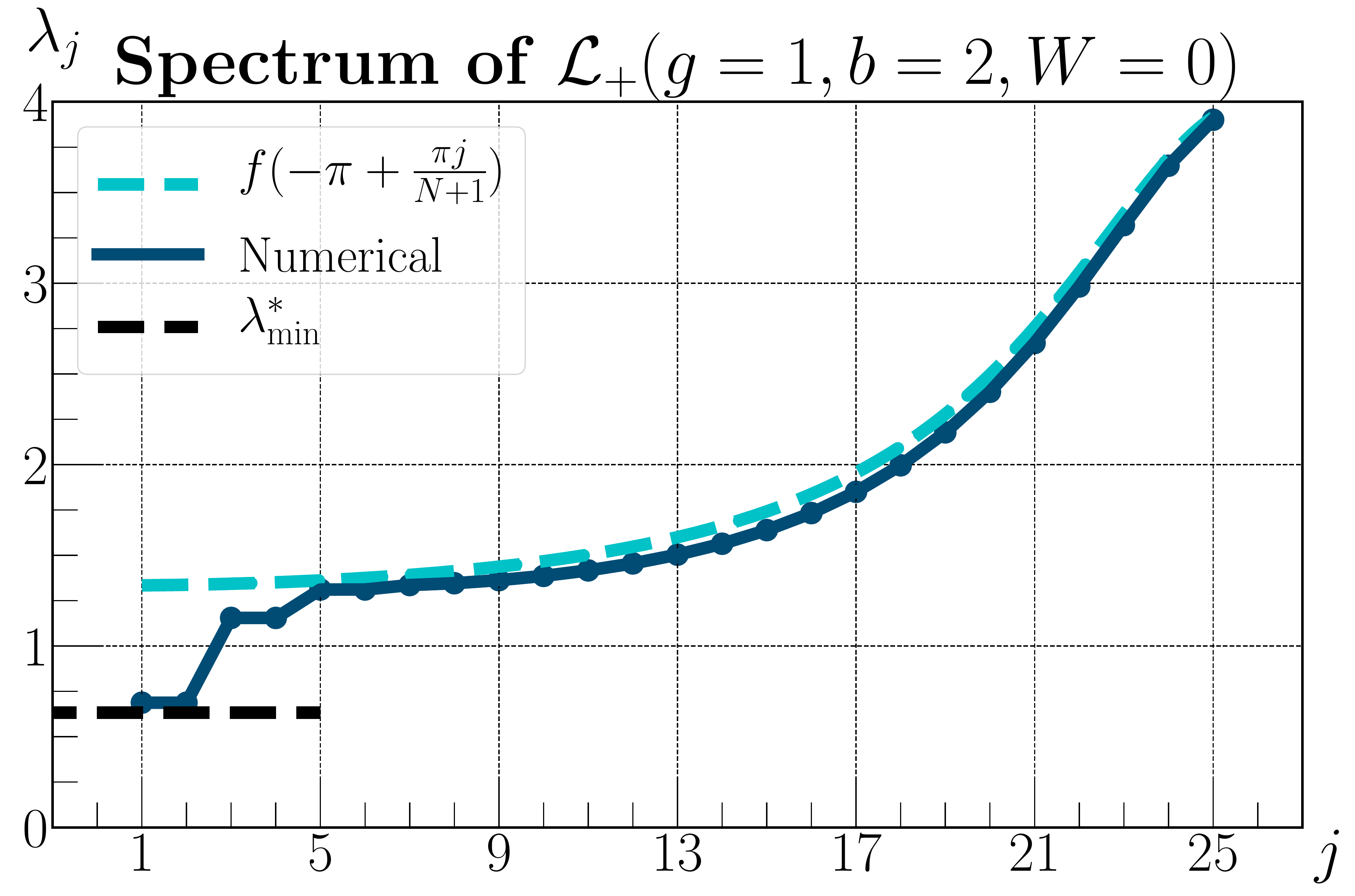}
     \caption{Eigenvalues of $\mathcal{L}_+(g = 1, b = 2, W = 0)$ with analytical predictions alongside.}
     \label{fig:L_+ Spectrum}
 \end{figure}

It is also worth noting that this technique is quite general and can be used to bound the spectra of the mass matrices considered in sections~\ref{sec:anderson-inspired} and~\ref{sec:nonlocaltheoryspacemodels}. These mass matrices are slightly more complicated, however, because we cannot simply write them as the sum of a Toeplitz matrix and a diagonal matrix with $\epsilon_j$'s as entries. Instead, it is useful to write the mass matrix as the sum of a Toeplitz part, $T$, a diagonal remainder part, $D$, and an $\epsilon$-dependent random part, $P$. For instance, one would decompose the mass matrix of $\mathcal{L}_+$ (Eq. \ref{eq:non-Local Lagrangian1}) as follows
\begin{equation}
 \mathcal{L}_+ \supset \vec{\pi}^T \cdot M_\pi \cdot \vec{\pi} \hspace{30pt} M_\pi = gT + gD + P   
\end{equation}
with 
\begin{equation}
\begin{split}
    T_{j,k} &= b^{-|j-k|}+\frac{3-b}{b-1} \delta_{j, k} \\
    D_{j,k} &= -(b^{1-j} + b^{j-N})\delta_{j,k} \\
    P_{j,k} &= \epsilon_j \delta_{j,k}
\end{split}
\end{equation}
Then, one could find the spectrum of $T$ by computing its symbol and incorporate the spectra of $D$ and $P$ (which are trivial to calculate because these matrices are diagonal) to discern the spectrum and bandwidth of $M$. This technique works for an arbitrary bisymmetric model too and is a powerful tool that can also be used to show that the smallest two eigenvalues of $\mathcal{L}_+$ are suppressed by an order one amount when compared to the rest of the spectrum. Concretely, to calculate the spectrum of $\mathcal{L}_+$, we start by computing the symbol of $T$, which gives a good estimate of most of the eigenvalues, as shown in Figure \ref{fig:L_+ Spectrum}. One can see that the two lightest eigenvalues are evidently smaller than the prediction of the symbol, however. To accurately approximate those two, we include the diagonal matrices $P$ and $D$ to first order in perturbation theory. This gives the following expected shifts to the lightest two eigenvalues
\begin{equation}
    \langle \Delta \lambda \rangle = \langle v^T \cdot P \cdot v \rangle +  v^T \cdot D \cdot v = \frac{W}{2} -\sum_{j,k} e^{(i-N)/L}(b^{1-j}+b^{j-N})\delta_{j,k}e^{(k-N)/L}, 
\end{equation}
 where we have used the fact that minimal mass eigenstates are localized around one of their end-sites. We plot the two lightest eigenvalues taking into account of the first order corrections, denoted $\lambda_{\text{min}}^*$, in Figure \ref{fig:L_+ Spectrum} as well. It is clear that our analytical predictions give a good estimate of the spectrum, and this perturbative technique offers a compelling explanation regarding why the lightest eigenvalues of $\mathcal{L}_+$ are relatively suppressed. Why aren't the other eigenvalues suppressed similarly? This is because the lightest eigenvectors have components that are localized around one of the end-site fields. Indeed, having larger components on end-sites means that $v^T \cdot D \cdot v$ is smaller because $D$ too has larger components near the edges of the diagonal that taper off exponentially as one proceeds to more central components. Because other eigenvectors corresponding to heavier eigenstates have more centralized support (to ensure orthogonality), they are less affected by $D$, to the first order in perturbation theory.
 
\section{Proof of Near Commutativity of $A$ and $BJ$}
\label{sec:near commutativity}

The goal of this appendix is to discuss the approximate commutativity of the matrices $A$ and $BJ$ defined in Eq.~\eqref{def of A, BR}, which is used in Sec.~\ref{sec:analytical considerations} for a semi-analytical understanding of the results. For simplicity of notations, we define the coupling, $g_{j} \equiv g/b^{j}$. 

We consider $N$ real scalars with $N$ being even. Similar arguments could be developed for odd $N$. Given Eq.~\eqref{eq:defofPC} and \eqref{def of A, BR}, we have $A$ and $BJ$ as $N/2 \times N/2$ matrices taking the forms:
\begin{equation}
    A = \begin{pmatrix}
    G_{N/2} & g_1 & \cdots & g_{N/2-1} \\
    g_1 & G_{N/2-1} & \cdots & g_{N/2 -2}\\
    \vdots & \ddots & \ddots & \vdots \\
    g_{N/2-1} & g_{N/2-2} & \cdots & G_{1}
    \end{pmatrix} \hspace{20pt}
    BJ = \begin{pmatrix}
    g_{1} & g_{2} & \cdots & g_{N/2} \\
    g_{2} & g_{3} & \cdots & g_{N/2 +1}\\
    \vdots & \ddots & \ddots & \vdots \\
    g_{N/2} & g_{N/2+1} & \cdots & g_{N-1}
    \end{pmatrix}, 
\end{equation}
where the diagonal entries $G_i$ are given by
\begin{equation}
    G_i = \sum_{k =1}^{N-i}g_k + \sum_{k=1}^{i-1}g_k. 
\end{equation}

To demonstrate the nearly commutativity of $A$ and $BJ$, it is useful to introduce several quantities. 
First, the Frobenius norm of a real matrix, $X$, is defined as:
\begin{equation}
    ||X||_F = \left(\sum_{i,j} X_{i,j}^2\right)^{1/2} = \sqrt{\text{Tr}(X^\dagger X)}. 
\end{equation}
We also define the sum of the modulus of the off-diagonal components squared as:
\begin{equation}
    \text{Off}(X) = \sum_{i \neq j} X_{i,j}^2.
\end{equation}
Lastly we define $\alpha(X,Y)$ as
\begin{equation}
\alpha(X,Y) = \text{inf} \{ \text{Off}(U^\dagger XU) + \text{Off}(U^\dagger YU)~|~ U \in U(N) \},
\end{equation}
where $\text{inf}$ indicates the greatest lower bound of the set. If $\alpha(X,Y)$ is small, there exists a unitary matrix that almost diagonalizes both matrices $X$ and $Y$. 

It has been shown that there exists a real valued function $\epsilon$ such that $\lim_{x \rightarrow 0} \epsilon(x) = 0$ and for all self-adjoint $N/2\times N/2$ matrices $X$ and $Y$ with unit Frobenius norm, 
\begin{equation}
    \alpha(X,Y) < \frac{N}{2} \epsilon(||[X,Y]||_F),
\end{equation}
with $\epsilon(x) = f(x^{-1}) x^{1/5}$ near the origin, where $f$ is a function independent of $N$ that grows slower than any power of $x$~\cite{bernstein1971almost}. In other words, if two self-adjoint matrices almost commute so that the right hand side of equation above is close to zero, they could almost be diagonalized at the same time and thus share similar eigenvectors.

Now, we present a proof that as long as $\sum_{i}g_{i} < \infty$, then the matrices $A$ and $BJ$ commute in the $N\rightarrow \infty$ limit.\footnote{Interestingly, $\sum_{i}g_{i} < \infty$ also leads to satisfying the Levitov criterion.} We introduce the normalized matrices $\bar{A}$ and $\overline{BJ}$ as
\begin{equation}
    \bar{A} \equiv \frac{A}{||A||_F}, \quad \overline{BJ} \equiv \frac{BJ}{||BJ||_F}. 
\end{equation}
It is straightfoward to show that 
\begin{equation}
    \text{min} \bigg\{ \frac{1}{||A||_F}, \frac{1}{||BJ||_F} \bigg\} \frac{\alpha(A, BJ)}{N} < \frac{\alpha(\bar{A}, \overline{BJ})}{N} < \frac{1}{2}\epsilon(||\left[\bar{A},\overline{BJ}\,\right]||_F). 
\end{equation}
In the large $N$ limit, it is clear that 
\begin{equation}
    ||A||_F \propto \sqrt{N}, \quad ||BJ||_F \propto \text{constant}. 
\end{equation}
Combining the equations above, we find that 
\begin{equation}
    \kappa \frac{\alpha(A,BJ)}{N^{3/2}} < \epsilon(|| \left[\bar{A},\overline{BJ} \, \right] ||_F), 
\end{equation}
where $\kappa$ is some real number independent of $N$. 
Next we evaluate the commutator on the right hand side in the equation above,
\begin{equation}
    ||\left[\bar{A},\overline{BJ}\, \right]||_F = \frac{||[A, BJ]||_F}{||A||_F ||BJ||_F} \propto \frac{||[A, BJ]||_F}{\sqrt{N}}. 
\end{equation}
Finally we want to prove that $||[A, BJ]||_F$ is bounded by some constant and doesn't grow with $N$. 
\begin{equation}
    \begin{split}
 ||[A, BJ] ||_F &= \bigg( \sum_i \sum_j \bigg[ \sum_k \left(A_{i,k}(BJ)_{k,j}- (BJ)_{i,k}A_{k,j}\right) \bigg]^2 \bigg)^{1/2} \\
        &= \bigg( \sum_i \sum_j \bigg[ \bigg(\sum_k A_{i,k}(BJ)_{k,j}\bigg)^2 -2\sum_k A_{i,k}(BJ)_{k,j}(BJ)_{i,k}A_{k,j} + \bigg(\sum_k (BJ)_{i,k}A_{k,j} \bigg)^2 \bigg] \bigg)^{1/2} \\
        & \leq  \bigg( \sum_i \sum_j \bigg[ \bigg(\sum_k A_{i,k}(BJ)_{k,j}\bigg)^2 + \bigg(\sum_k (BJ)_{i,k}A_{k,j} \bigg)^2 \bigg] \bigg)^{1/2} \\
        & = \bigg( 2  \sum_i \sum_j \bigg[ \sum_k A_{i,k}(BJ)_{k,j}\bigg]^2  \bigg)^{1/2} \\
    \end{split}
\end{equation}
To get the third line, we remove the middle term which is always negative in the models we consider as all elements of $A$ and $BJ$ are positive. It can be shown that this term is also bounded by a constant. We then insert the expressions for the matrix elements $(BJ)_{i,j} = g_{i+j-1}$, $A_{i,j} = g_{|i-j|}$ for $i \neq j$ and $A_{i,i} = G_{N/2 + 1 - i} \leq g_0$, where $g_j< g_0 \equiv \text{max}\{G_j~|~ j \in \mathbb{N} \} < \infty$. This gives
\begin{equation}
    \begin{split}
        ||[A, BJ]||_F &\leq \left( 2 \sum_i \sum_j \bigg[ \sum_k g_{|i-k|} g_{k+j-1} \bigg]^2 \right)^{1/2} \\
        & \leq \bigg(2 \sum_j \bigg[\sum_k g_{k+j-1} \sum_i g_{|i-k|}\bigg]^2 \bigg)^{1/2} \\
        & = \bigg(2 \sum_j \bigg( \sum_k g_{k+j-1} \bigg[ \sum_{i=0}^{k-1} g_{k-i} + \sum_{i=k}^\infty g_{i-k} \bigg] \bigg)^2 \bigg)^{1/2} \\
        &= \bigg(2 \sum_j \bigg( \sum_k g_{k+j-1} \bigg[ \sum_{i=1}^{k} g_{i} + \sum_{i=0}^\infty g_{i} \bigg] \bigg)^2 \bigg)^{1/2} \\
        &\leq \bigg(8 \sum_j \bigg( \sum_k g_{k+j-1} \sum_{i=0}^{\infty} g_{i} \bigg)^2 \bigg)^{1/2} \\
        &= \bigg(8 \bigg( \sum_i g_i \bigg)^2 \sum_j \bigg(\sum_k g_{k+j-1} \bigg)^2 \bigg)^{1/2} = \text{finite},
    \end{split}
\end{equation}
where in the last line, we use $\sum_j g_j$ and $\sum_j (\sum_kg_{k+j-1})^2$ are finite.

In summary, we have demonstrated that 
\begin{equation}
    \kappa \frac{\alpha(A,BJ)}{N^{3/2}} < \epsilon\Big(\frac{\eta}{N^{1/2}}\Big), 
\end{equation}
where $\kappa, \eta$ are real constants. Denoting the average value of the squares of the off-diagonal components of the matrix $X$ as $\langle X_{i,j} \rangle|_{\text{off}}$, we have $\alpha(A, BJ) = N(N-1)(\langle U^\dagger A U  \rangle|_{\text{off}} + \langle U^\dagger (BJ) U \rangle|_{\text{off}})$. Given the behavior of $\epsilon(x)$ near the origin in the $N \rightarrow \infty$ limit, we have
\begin{equation}
    \langle U^\dagger A U  \rangle|_{\text{off}} + \langle U^\dagger (BJ) U  \rangle|_{\text{off}} \sim N^{-3/5}. 
\end{equation}
Therefore, in this limit, the matrices $A$ and $BJ$ can be simultaneously diagonalized, and, accordingly, components of their eigenvectors converge. Indeed, the fraction of components of $U^\dagger A U$ (and $U^\dagger(BJ) U$) that are non-zero (given by $N^{-3/5}$) tends to zero demonstrating that eigenvectors of $A$ and $BJ$ agree except perhaps at a finite fraction of sites.

\bibliography{ref}
\bibliographystyle{utphys.bst}
\end{document}